\documentclass[aps,twocolumn,pra,superscriptaddress,10pt]{revtex4-2}

\usepackage[T1]{fontenc}
\usepackage{dsfont}
\usepackage{textgreek}
\usepackage{amsfonts}
\usepackage{amsmath}
\usepackage{bm}
\usepackage{bbold}
\usepackage{amssymb}
\newcommand{\Mod}[1]{\ (\mathrm{mod}\ #1)}
\usepackage{graphicx}
\usepackage{braket}

\usepackage[colorlinks=true,urlcolor=blue,citecolor=blue,linkcolor=blue]{hyperref}
\DeclareGraphicsExtensions{.eps,.ps,.pdf}
\newtheorem{identity}{Identity}
\newcommand{\diff}{\mathop{}\!\mathrm{d}}

\newcommand{\si}{\sigma}

\newcommand{\BM}{\begin{displaymath}}

\newcommand{\EM}{\end{displaymath}}
\newcommand{\IMA}{\textrm{im}}

\newcommand{\ie}{\hbox{\em i.e.{}}}

\def \Tr {\text{Tr}}

\newcommand{\ep}{e_{\text{p}}}
\newcommand{\TT}{\mathcal{T}}
\newcommand{\NN}{\mathcal{N}}
\newcommand{\MM}{\mathcal{M}}
\newcommand{\PP}{\mathcal{P}}
\newcommand{\UU}{\mathcal{U}}
 \newcommand{\ED}{\mathcal{D}}
\newcommand{\MUU}{\widehat{\UU}}

\newcommand{\Hs}{\mathcal{H}}

\newcommand{\HSs}{\mathcal{HS}}

\makeatletter
\g@addto@macro\bfseries{\boldmath}
\makeatother

\begin{document}

\widetext

\title{Entangling power of symmetric multiqubit systems: a geometrical approach} 

\author{Eduardo Serrano-Ens\'astiga}
\email{ed.ensastiga@uliege.be}
\affiliation{Institut de Physique Nucléaire, Atomique et de Spectroscopie, CESAM, University of Liège, B-4000 Liège, Belgium}

\author{Diego Morachis Galindo}
\affiliation{Institut de Physique Nucléaire, Atomique et de Spectroscopie, CESAM, University of Liège, B-4000 Liège, Belgium}
\affiliation{Departamento de F\'isica, Centro de Nanociencias y Nanotecnolog\'ia, Universidad Nacional Aut\'onoma de M\'exico\\
Apartado Postal 14, 22800, Ensenada, Baja California, M\'exico}

\author{Jesús A. Maytorena}
\affiliation{Departamento de F\'isica, Centro de Nanociencias y Nanotecnolog\'ia, Universidad Nacional Aut\'onoma de M\'exico\\
Apartado Postal 14, 22800, Ensenada, Baja California, M\'exico}

\author{Chryssomalis Chryssomalakos}
\affiliation{Instituto de Ciencias Nucleares, Universidad Nacional Autónoma de México PO Box 70-543, 04510, CDMX, México}

\date{\today}
\begin{abstract}
Unitary gates with high entangling capabilities are relevant for several quantum-enhanced technologies. For symmetric multiqubit systems, such as spin states or bosonic systems, the particle exchange symmetry restricts these gates and also the set of not-entangled states. In this work, we analyze the entangling power of unitary gates in these systems by reformulating it as an inner product between vectors with components given by SU$(2)$ invariants. For small number of qubits, this approach allows us to study analytically the entangling power including the detection of the unitary gate that maximizes it. We observe that extremal unitary gates exhibit entanglement distributions with high rotational symmetry, same that are linked to a convex combination of Husimi functions of certain states. Furthermore, we explore the connection between entangling power and the Schmidt numbers admissible in some quantum state subspaces. Thus, the geometrical approach presented here suggests new paths for studying entangling power linked to other concepts in quantum information theory.
\end{abstract}
\maketitle
\section{Introduction}
Entanglement is a foundational concept in quantum theory and a vital resource for quantum technologies such as quantum computing, cryptography, metrology and simulation~\cite{RevModPhys.81.865,Nie.Chu:11,Ben.Zyc:17}. In multipartite quantum systems, it is generated through nonlocal unitary transformations, since local operations cannot alter the entanglement of a state~\cite{RevModPhys.81.865}. It is therefore natural to investigate the entangling capacity of nonlocal unitary gates, and how to create highly entangled states via the evolution of nonlocal Hamiltonians or pulse sequences in physical systems~\cite{PhysRevLett.87.137901,PhysRevLett.90.047904,Xiong_2012,PhysRevA.63.062309,PhysRevA.67.042313,Kalsi_2022,PhysRevApplied.19.044094,PhysRevA.81.062346}.
On the theoretical side, several concepts have been proposed to assess the capability of unitary gates in generating quantum resources such as \emph{gate typicality}, \emph{disentangling power}, or \emph{perfect entanglers}~\cite{PhysRevA.62.030301,PhysRevA.67.052301,PhysRevA.95.040302,CLARISSE2007400}. One of the most intuitively appealing quantities is the \emph{entangling power} of a unitary gate~\cite{PhysRevA.62.030301} which is defined as the average entanglement generated by a unitary gate over the set of separable (not-entangled) states. Extensive studies of the entangling power have been carried out in bipartite systems~\cite{PhysRevA.63.040304,PhysRevA.70.052313,PhysRevA.79.052339,Shen_2018,PhysRevA.67.042323,YANG20084369,PhysRevA.87.022111,Chen_2019} with generalizations to two-qubit global noise channels~\cite{Kong_2024} and
multipartite systems~\cite{PhysRevA.69.052330,Linowski_2020}. The entangling power of a unitary operator is also connected to its entanglement in its own Schmidt decomposition~\cite{PhysRevA.63.040304,PhysRevA.66.044303} or to invariant quantities under local transformations~\cite{makhlin2002nonlocal,PhysRevA.67.042313,PhysRevA.82.034301}. Furthermore, connections have been identified between highly entangling unitary gates and Quantum Error-Correcting Codes~\cite{PhysRevA.69.052330}, absolutely maximally entangled (AME) states~\cite{Linowski_2020}, unitary operators invariant under local actions of diagonal unitary and orthogonal groups~\cite{Singh_2022}, and quantum versions of chaotic systems~\cite{A_J_Scott_2003,PhysRevB.98.174304,PhysRevResearch.3.043034}. 

When a multiqubit system is restricted to its symmetric subspace, the set of product states is significantly reduced to the set of Spin-Coherent (SC) states, which form a 2-sphere within the Hilbert space of quantum states~\cite{Chr.Guz.Ser:18,Ben.Zyc:17}. Additionally, the local unitary transformations are limited to the global rotations, specifically global SU$(2)$ transformations generated by the angular momentum operators. Symmetric multiqubit states arise in bosonic systems, such as two-mode multiphotonic systems or spin-$j$ states. Moreover, the symmetric subspace of $N$ qubit states is equivalent to the Hilbert space of spin-$j$ ($j=N/2$) states, which makes the study of the entangling power of single spin states relevant in distinct physical platforms.

For the symmetric two-qubit case, which can be thought of as a spin-1 system, the entangling power of $3 \times 3$ unitary matrices has been studied in Ref.~\cite{PhysRevA.105.012601}. One key mathematical feature that arises in the study of those unitaries is the Cartan decomposition, which factorizes any quantum gate as a non-local operation multiplied on each side by local $\mathrm{SU}(2)$ transformations~\cite{10.1063/1.532618}. The entangling power of the unitary is encoded into the non-local factor, which is obtained by exponentiating the maximally commuting (Cartan) subalgebra of $\mathfrak{su}(3)$. This allows to represent the unitaries with equivalent entangling properties in an euclidean two-dimensional space~\cite{PhysRevA.105.012601}. For symmetric $N$-qubit systems with $N >2$, the Cartan decomposition would not represent such a significant advantage, since the number of relevant parameters grows quadratically with the dimension of the system. 

In this work, we study the entangling power of unitary transformations acting in symmetric multiqubit systems. We reformulate its original expression in terms of $\mathrm{SU}(2)$ invariant quantities of the unitary operators. This geometrical approach connects the entangling power to other well-known quantities in quantum information theory. 

The structure of the paper is as follows: Sec.~\ref{Sec.Theory} considers symmetric multiqubit states and reviews, in this context, the formal definitions of the entropy of entanglement and of the entangling power --- we derive the above mentioned geometric expression of the latter in Sec.~\ref{Sec.Reformulation}. We then calculate the entangling power and the entanglement distribution on the sphere for systems with a small number of qubits in Sec.~\ref{Sec.Examples} and \ref{Sec.Husimi}, respectively, including the search of unitary gates with the best entangling capabilities. Sec.~\ref{Sec.Average} contains the calculation of the average entangling power over all unitary gates. An interesting connection between entangling power and Schmidt numbers is explored in \ref{Sec.Schmidt}. We give some last remarks in Sec.~\ref{Sec.Conclusions}.
\section{Basic concepts}
\label{Sec.Theory}
\subsection{Equivalence between symmetric multiqubit states and spin-\texorpdfstring{$j$}{Lg} states}
We start this section by briefly describing the equivalence between the symmetric sector of $N$ qubits and a spin-$j=N/2$ state (see Ref.~\cite{Den.Mar:22} for a detailed discussion of this equivalence). The symmetric sector of the $N$-qubit system is spanned, for instance, by the Dicke states $\ket{D^{(N)}_k}$
\begin{equation}
\ket{D^{(N)}_k} = K \sum_{\Pi } \Pi \left( \underbrace{\ket{+} \otimes \dots \otimes \ket{+}}_{N-k} \otimes \underbrace{\ket{-} \otimes \dots \otimes \ket{-}}_{k} \right) ,
\end{equation}
where $K$ is a normalization factor and the sum runs over all the permutations of the qubits. If we consider that the qubits are spins $1/2$, the states $\ket{D^{(N)}_k}$ are eigenvectors of the angular momentum operators $J_z$ and $\mathbf{J}^2=J_x^2 + J_y^2+J_z^2$
\begin{equation}
\begin{aligned}
\mathbf{J}^2\ket{D^{(N)}_k} &=j(j+1) \ket{D^{(N)}_k} , 
\\
J_z \ket{D^{(N)}_k} &= m \ket{D^{(N)}_k} .
\end{aligned}
\end{equation}
with $m=j-k$.
Thus, the span of the Dicke states constitute a spin-$j$ Hilbert space, denoted as $\Hs^{(j)}$, and where we can identify $\ket{ D_k^{(N)} }= \ket{j,m}$. This space constitutes a spinor system because its elements transform under rotations with respect to the spin-$j$ irreducible representation (irrep) of $\mathrm{SU}(2)$. From now on, we mostly work in the language of spins where the SU$(2)$ irreps appear naturally.
\subsection{Bipartite entanglement and entangling power}
We consider a single spin-$j$ (symmetric $N=2j$-qubits) system with Hilbert space  $\Hs^{(j)}$ and the Hilbert-Schmidt (HS) space of bounded operators $\HSs(\Hs^{(j)})$ acting on $\Hs^{(j)}$. In this framework, we consider a bipartition of these $2j$ qubits $\Hs^{(j)} \subset \Hs_A \otimes \Hs_B$ and assume, without loss of generality, that $\dim (\Hs_A ) \leqslant \dim (\Hs_B) $. Moreover, since any subsystem of a symmetric state is also symmetric, we can take as subsystems $\Hs_A \cong \Hs^{(q/2)}$ and $\Hs_B \cong \Hs^{(j-q/2)}$. In other words, the bipartition involves states of spins $q/2$ and $j-q/2$, respectively. 
The entanglement of a bipartite state $\ket{\Psi} \in \Hs_A \otimes \Hs_B$ can be quantified by the (normalized) linear entanglement entropy~\cite{PhysRevA.62.030301}
\begin{equation}
\label{Eq.Ent.generalAB}
    E (\ket{\Psi}) \equiv \frac{d}{d-1}\left[ 1- \Tr \left( \rho_A^{ 2} \right) \right]  ,
\end{equation}
where $E (\ket{\Psi}) \in [0,1]$, $\rho_{A}= \Tr_B (\ket{\Psi}\bra{\Psi})$ is the reduced mixed state after tracing out the subsystem $B$, and $d= \dim \left(  \Hs_A \right)$. For a general bipartite pure state with Schmidt decomposition
\begin{equation}
    \ket{\Psi} = \sum_{k=1}^d \sqrt{\Gamma}_k
\ket{\psi_{A,k}}\otimes \ket{\psi_{B,k}} ,
\end{equation}
where $\{ \ket{\psi_{A,k}} \}_k \subset \Hs_A$ and $\{ \ket{\psi_{B,k}}\}_k \subset \Hs_B$ are orthogonal sets of states and $\sum_k \Gamma_k=1$, we have
\begin{equation}
        E (\ket{\Psi}) = \frac{d}{d-1}\left[ 1- \sum_{k=1}^d \Gamma_k^2 \right]  .
\end{equation}
In particular, $E(\ket{\Psi})=0$ for bipartite product states $\ket{\Psi} = \ket{\psi_A} \otimes \ket{\psi_B}$. On the other hand, the maximally entangled states have Schmidt numbers $\Gamma_k= 1/\sqrt{d}$ and $E(\ket{\Psi})=1$. Each bipartition with $q \leq \lfloor j \rfloor$ defines a different measure of entanglement $E_q$. This because $\Tr(\rho_A^2) = \Tr(\rho_B^2)$, and thus measures of entanglement $E_q$ for $q> \lfloor j \rfloor$ are linear combinations of those of lower $q$ values. 

The product (separable) states in $\Hs^{(j)}$, for any bipartition, are the spin-coherent (SC) states $\ket{j,\mathbf{n}} \equiv D^{(j)}(\mathbf{n}) \ket{j,j}$ which constitute a 2-sphere in $\Hs^{(j)}$~\cite{Chr.Guz.Ser:18}. A possible parametrization of them is given by the rotations $D^{(j)}(\mathbf{n}) = D^{(j)}(0,\theta ,\phi)$ which align the $\mathbf{z}$ axis to the direction $\mathbf{n}$ with spherical angles $(\theta, \phi)$. $D^{(j)}(\alpha, \beta ,\gamma)$ denotes the irreducible representation $j$ ($j$-irrep) of a rotation matrix in the Euler angle parametrization~\cite{Var.Mos.Khe:88}. The \emph{entangling power} of a quantum unitary gate  $U \in \HSs (\Hs^{(j)})$, with respect to $E_q$, is defined as the average entanglement produced by $U$ acting on the SC states~\cite{PhysRevA.62.030301},
\begin{equation}
\label{Eq.entangling.power}
\ep(E_q,U) \equiv \overline{E_q(U \ket{j,\mathbf{n}})} = \frac{1}{4\pi} \int_{S^2} E_q(U \ket{j,\mathbf{n}}) \diff \mathbf{n}  .
\end{equation}
It is easily seen that 
\begin{equation}
\label{Eq.invariance.ep}
    \ep \big(E_q,R_1 U R_2 \big) = \ep(E_q,U) ,
\end{equation}
for all $q$, where $R_{1,2}$ are matrices representing arbitrary $\mathrm{SU}(2)$ elements.
 If $U$ itself is an $\mathrm{SU}(2)$ element, then $\ep (E_q , U) = 0$ for all $q$.

Lastly, and since we will use for our main results, we specify the coupled basis of two spins
\begin{equation}
\label{Eq.coupled.basis}
 \ket{j_1,j_2,L,M} \equiv \hspace{-0.2cm}
 \sum_{m_2 =-j_2}^{j_2} \sum_{m_1 =-j_1}^{j_1} \hspace{-0.2cm} C_{j_1 m_1 j_2 m_2}^{LM}  \ket{j_1,m_1} \ket{ j_2 , m_2} ,
\end{equation}
with $\ket{j_1,m_1} \ket{ j_2 , m_2}= \ket{j_1,m_1} \otimes \ket{ j_2 , m_2}$, and $C_{j_1 m_1 j_2 m_2}^{LM}$ denoting the Clebsch-Gordan coefficients.
\subsection{Case \texorpdfstring{$j=1$ (N=2)}{Lg}}
\begin{figure}[t!]
\includegraphics[angle=0,width=.40\textwidth]{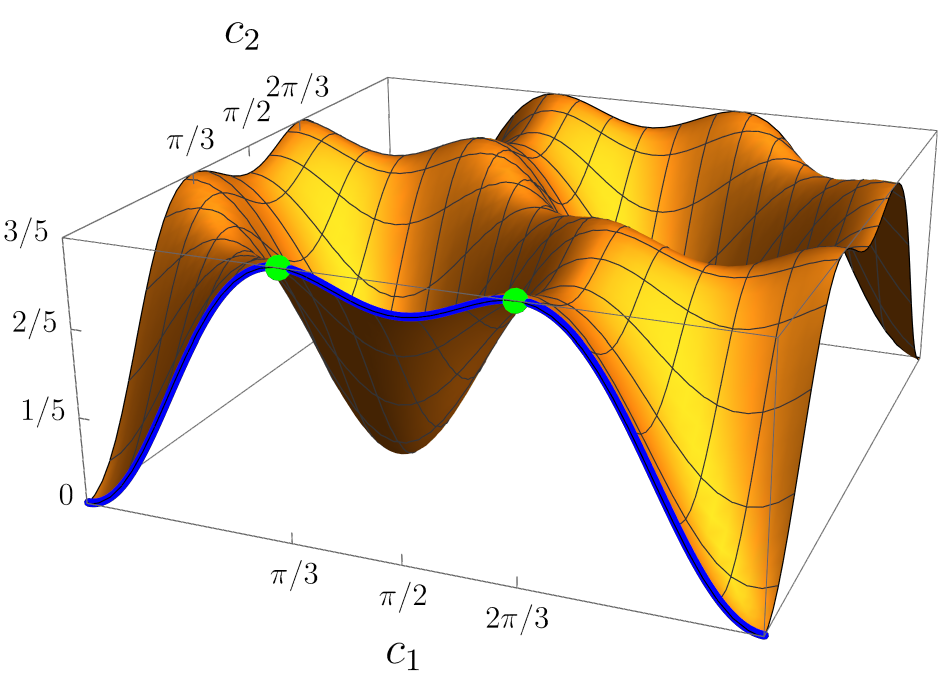}
\caption{
Plot of the function $\ep(E_1,U)$ given in Eq.~\eqref{epc1c2}. One can clearly access a maximum with $c_2=0$ (blue curve). Two particular maxima are shown (green points); the one on the right, corresponding to $c_1=2\pi/3$, gives the $U_0$ in~(\ref{U0M}).
}
\label{Fig.U0.spin1}
\end{figure}

The entangling power $\ep(E_q,U)$ for $j=1$ and  $q=1$ can be written as~\cite{PhysRevA.105.012601}
\begin{equation}
\label{Eq.ep.spin1}
\ep (E_1 , U)= \frac{3}{5} \left(1-  \frac{1}{9}  \left| \Tr \left( m \right) \right|^2 \right) , 
\end{equation}
where 
\begin{equation}
\label{Eq.Matri.m}
    \Tr (m) = \Tr (U_B^T U_B) ,
\end{equation}
and $U_B$ is the unitary matrix $U$ transformed in the  symmetric Bell states basis,
$U_B= Q^{\dagger} U Q$ \footnote{Here, we consider the $Q$ matrix of \cite{PhysRevA.105.012601} in the symmetric subspace of $\Hs_{1/2}^{\otimes 2}$.} with
\begin{equation}
Q=\frac{1}{\sqrt{2}} \left(
\begin{array}{ccc}
 1 & 0 & i \\
 0 & i \sqrt{2} & 0 \\
 1 & 0 & -i \\
\end{array}
\right) .
\end{equation}
After some algebra, we can also write Eq.~\eqref{Eq.Matri.m} as
\begin{equation}
\label{m.exp1}
\Tr (m) = \Tr \left( \Phi  U^T \Phi  U \right)  , \quad \text{with } 
\Phi=\left(
\begin{array}{ccc}
0 & 0 & -1
\\
0 & 1 & 0
\\ 
-1 & 0 & 0
\end{array}
\right)
.
\end{equation}
The unitary transformations of spin-1 states can be parametrized using the Cartan decomposition~\cite{PhysRevA.105.012601,10.1063/1.532618}, with the SU$(2)$ subgroup generated by the $j=1$ angular momentum operators. A possible parametrization is $U=R_1 A R_2$ with $R_1, R_2 \in \mathrm{SU}(2)$ and 
\begin{equation}
\label{Ec:Anonlocal}
A= \left(
\begin{array}{c c c}
\frac{\lambda_1 + \lambda_3}{2} & 0 & \frac{\lambda_1 - \lambda_3}{2}
\\
0 & \lambda_2 & 0
\\
\frac{\lambda_1 - \lambda_3}{2} & 0 & \frac{\lambda_1 + \lambda_3}{2}
\end{array}
\right) ,
%\label{Ec:Anonlocal}
\end{equation}
with 
\begin{equation}
\begin{aligned}
\lambda_1 = & e^{\frac{i}{2} \left( -c_1 + c_2 + c_3 \right)}  , \quad
\\
\lambda_2 = & e^{\frac{i}{2} \left( c_1 + c_2 - c_3 \right)}  , \quad
\\
\lambda_3 = & e^{\frac{i}{2} \left(  c_1 - c_2 + c_3 \right)}  ,
\end{aligned}
\end{equation}
and where the real parameters $c_k$ fulfill $c_1+c_2+c_3=0$~\cite{PhysRevA.105.012601}. Given that the entangling power is invariant under left and right rotations (see Eq.~\eqref{Eq.invariance.ep}), $\ep(E_1,U)$ only depends on the $c_i$'s,
\begin{equation}
\label{epc1c2}
\ep( E_1 , U)= 
\frac{4}{15} \left( \sin^2 c_{12}+\sin^2c_{13} +\sin^2 c_{23} \right)  ,
\end{equation}
with $c_{ij}=c_i - c_j$. We plot $\ep(E_1 ,U)$ as a function of $c_1$ and $c_2$ in Fig.~\ref{Fig.U0.spin1}. For $c_2=0$, we observe two unitary matrices attaining the maximum at $c_1 = \pi/3 $ and $2\pi/3$, respectively. The latter solution has corresponding unitary matrix (with $R_{1,2}=I$)
\begin{equation}
\label{U0M}
U_0=
\left(
\begin{array}{ccc}
\frac{1}{2}(\omega+1) & 0 & \frac{1}{2}(\omega-1)
\\
0 & \omega^{-1} & 0
\\
\frac{1}{2}(\omega-1) & 0 & \frac{1}{2}(\omega+1)
\end{array}
\right)
 ,
\end{equation}
where $\omega=e^{-i \pi/3}$ is a cubic root of $-1$ and $\ep(E_1 ,U_0)=3/5$. One can find two rotations to construct another unitary gate $U_0' =R_1 U_0 R_2 $ with the same entangling power as $U_0$ but consisting simply of a permutation matrix
\begin{equation}
\label{Eq.j1.U0p}
    U_0' = \left( 
    \begin{array}{ccc}
         1 & 0 & 0  \\
         0 & 0 & 1 \\
         0 & 1 & 0
    \end{array}
    \right) .
\end{equation}
%
%%%%%%% 
%
%%%%%%%%%%%%%%%%%% BEGIN FIGURE
%
%
%%%%%%%%%%%%%%%%% END FIGURE
%
%
%
\section{Entangling power for symmetric \texorpdfstring{$N=2j$}{Lg} qubits}
\label{Sec.Reformulation}
Our first result is the reformulation of $E_q(U\ket{j,\mathbf{n}})$ in the form
\begin{equation}
\label{Eq.Ent.q.first}
\begin{aligned}    
E_q(U\ket{j,\mathbf{n}}) = & 1- \bra{2j,\mathbf{n}} \UU^{\dagger} \MM_q \UU \ket{2j,\mathbf{n}} ,
\end{aligned}
\end{equation}
where $\UU \equiv U \otimes U$ and
\begin{equation}
\label{Eq.Exp.Mq.PL}
\MM_q  \equiv \frac{q+1}{q} \sum_{L=0}^{2j}  (-1)^{2j+L} \chi(q,j,L)
 \PP_L,
\end{equation}
with $\UU, \MM_q \in \HSs(\Hs^{(j)\otimes 2} )$ and
\begin{equation}
    \chi(q,j,L) \equiv \sum_{\si=1}^{q}
(2\si+1) \frac{    
\left\{ 
\begin{array}{ccc}
    q/2 & q/2 & q  \\
    q/2 & q/2 & \si
\end{array} 
\right\}
\left\{ 
\begin{array}{ccc}
    j & j & L  \\
    j & j & \si
\end{array} 
\right\}
}{\left\{ 
\begin{array}{ccc}
    j & j & 2j  \\
    j & j & \si
\end{array} 
\right\} 
}
\end{equation}
and
\begin{equation}
\PP_L \equiv \sum_{M=-L}^L \ket{j,j,L,M}\bra{j,j,L,M} .
\end{equation}
Here, the curly bracket represents the Wigner 6-j symbol~\cite{Var.Mos.Khe:88}.
$\PP_L$ is the projector operator in the subspace of $\Hs^{(j)\otimes 2}$ defined with the states of coupled basis~\eqref{Eq.coupled.basis} with total angular momentum $L$. The derivation of Eq.~\eqref{Eq.Ent.q.first} is given in Appendix~\ref{Appendix.Main.Equation}. We now use the resolution of unity of the SC states~\cite{PhysRevA.24.2889},
\begin{equation}
    \frac{1}{4\pi} \int \ket{2j , \mathbf{n}} \bra{2j , \mathbf{n}} \diff \mathbf{n} = \frac{\PP_{2j}}{4j+1} \equiv \NN  ,
\end{equation}
 to calculate $\ep$, yielding 
\begin{equation}
\label{epU.exp.with.T}
\begin{aligned}    
\ep(E_q , U) =  &
1- 
 \Tr \left( \UU \NN  \UU^{\dagger} \MM_q \right)  .
\end{aligned}
\end{equation}
The fact that both operators $\NN$ and $\MM_q$ are linear combinations of $\PP_{L}$'s, which are orthogonal among themselves, $\Tr (\PP_L \PP_K) = (2L+1)\delta_{LK}$, suggests a further reformulation of the $\ep$. Indeed, to an operator $V \in \HSs(\Hs^{(j)\otimes 2})$, we associate the $(2j+1)$-dimensional vector $\overrightarrow{V}$ with $\mathrm{SU}(2)$-invariant components,
\begin{equation}
\label{Eq.Vec.SU2}
\overrightarrow{V}  = \Bigg( \Tr \left( V \tilde{\PP}_0 \right) , \dots , \Tr \left( V \tilde{\PP}_{L} \right) , \dots , \Tr \left( V \tilde{\PP}_{2j} \right) \Bigg) ,
\end{equation}
where $\tilde{\PP}_{L}= \PP_{L} / \sqrt{2L+1}$. We obtain that
\begin{equation}
\label{Eq.Trace.InnerP}
    \left\langle \MUU \overrightarrow{  \NN   }  , \overrightarrow{\MM}_q  \right\rangle
    =  \Tr \left( \UU \NN  \UU^{\dagger} \MM_q \right) ,
\end{equation}
where $\langle \cdot, \cdot \rangle$ is the euclidean inner product, 
\begin{equation}
\label{Eq.Inner.Product}
    \left\langle \overrightarrow{V}, \overrightarrow{W} \right\rangle \equiv \sum_{k=0}^{2j} V_k W_k  ,
\end{equation}
and $\MUU$ is a $(2j+1)\times (2j+1)$ real matrix with entries
\begin{equation}
\label{Eq.Def.MatrixU}
\left( \MUU \right)_{mn} = \Tr \left( \UU \tilde{\PP}_m \UU^{\dagger} \tilde{\PP}_n \right) .
\end{equation}
Therefore, the entangling power reads
\begin{equation}
\label{Eq.EP.vec}
    \ep (E_q ,U)= 1-  \left\langle \MUU  \overrightarrow{  \NN   }  , \overrightarrow{\MM}_q \right\rangle
    .
\end{equation}
From Eqs.~\eqref{Eq.Inner.Product} and~\eqref{Eq.Def.MatrixU}, we deduce that 
$\big\langle \MUU \overrightarrow{\NN} , \overrightarrow{\MM}_q  \big\rangle
= \big\langle  \overrightarrow{\NN}   , \MUU^{T} \overrightarrow{\MM}_q  \big\rangle$. To summarize, we have expressed $\ep(E_q,U)$ in terms of the euclidean inner product of two vectors, with components given by SU$(2)$-invariant quantities of the operators $\NN$ and $\MM_q$, one of them transformed by the unitary transformation $\UU$.
A similar formula for $\ep$, in terms of operators associated to the multipole operators~\cite{Var.Mos.Khe:88}, is given in Appendix~\ref{App.T.Basis}.

The projectors $\PP_L$ change by a $(-1)^{2j+L}$ sign under particle exchange in $\Hs^{(j)\otimes 2}$. Since the unitary operators $\UU = U \otimes U$ preserve this symmetry, $\Tr (\UU \PP_{2j} \UU^{\dagger} \PP_{L} )$ vanishes unless $L \equiv 2j \Mod{2}$. Thus, the vector $\overrightarrow{\NN}= (0,0,\dots , 1/\sqrt{4j+1} )$ after a $\UU$ transformation has components
\begin{equation}
\label{Eq.comp.P}
    \begin{array}{l}
   \Bigg( 0, \Tr \left( \NN \tilde{\PP}_{1} \right) , 0 , \Tr \left( \NN \tilde{\PP}_{3} \right) , 0 , \dots  , \Tr \left( \NN \tilde{\PP}_{2j} \right) \Bigg) ,
   \\[12pt]
   \Bigg( \Tr \left( \NN \tilde{\PP}_{0} \right) , 0 , \Tr \left( \NN \tilde{\PP}_{2} \right) , 0 , \dots  , \Tr \left( \NN \tilde{\PP}_{2j} \right) \Bigg) ,
    \end{array}
\end{equation}
for $2j$ odd or even, respectively.
On the other hand, $\overrightarrow{\MM}_q$ has components for $L$ both odd and even. As an example, the components of $\overrightarrow{\MM}_1$ are
\begin{equation}
    \left( \overrightarrow{\MM}_1 \right)_L = \frac{L(L+1)-2j(j+1)}{2j^2} .
\end{equation}
Nevertheless, the relevant components of $\MM_q$ for $\ep$ are the ones in common with $\MUU \overrightarrow{N}$~\eqref{Eq.comp.P}, \ie, with $2j\equiv L \Mod{2}$.
 These components of $\overrightarrow{\NN}$ and $\overrightarrow{\MM}_q$ lie in a hyperplane after any unitary transformation $\UU=U\otimes U$
(see \textbf{Identity~\ref{Identity.3}} of Appendix~\ref{App.useful.id})
\begin{equation}
\label{Eq.Hyperplane.P}
\begin{aligned}    
    \sum_{\substack{L=0 \\ 2j\equiv L \Mod{2}} }^{2j}
    \hspace{-0.5cm} \sqrt{2L+1} \left(\MUU \overrightarrow{\NN} \right)_L
    & = \hspace{-0.3cm}
    \sum_{\substack{L=0 \\ 2j\equiv L \Mod{2}} }^{2j} \hspace{-0.5cm}  \Tr \left( \UU \PP_{L} \UU^{\dagger} \NN \right) 
    = 1  ,
    \\
    \sum_{\substack{L=0 \\ 2j\equiv L \Mod{2}} }^{2j} 
    \hspace{-0.5cm} \sqrt{2L+1} \left(\MUU \overrightarrow{\MM}_q \right)_L 
    & =
    \frac{ (j+1)(2j+1)}{ 2j+1-q }     
     .
\end{aligned}
\end{equation}
The components of $\overrightarrow{\MM}_q$ for several spin values are shown in Table~\ref{Table.ep} --- they satisfy the inequalities
\begin{equation}
0 \leqslant \Tr \left( \UU \PP_{L} \UU^{\dagger} \PP_K \right)  
\leqslant
\min \left( 2L+1 , 2K+1 \right)
,
\label{ineqPLK}
\end{equation}
for $L\equiv K \Mod{2}$. This becomes evident we take into account that the operators $\PP_{L}$ and $\PP_K $ project onto subspaces of dimensions $2L+1$ and $2K+1$, respectively. The unitary operator transforms this subspace while preserving its dimension. The trace in~(\ref{ineqPLK}) then is just the projection of one of the subspaces, transformed by $\UU$, onto the other.  These inequalities can be used to find bounds for the components of the vector  
\begin{equation}
\label{Eq.bounds.components.P2j}
p_L \equiv \left( \MUU \overrightarrow{\NN} \right)_L
, \quad \quad
0 \leqslant p_L \leqslant \frac{\sqrt{2L+1}}{4j+1}  .
\end{equation}
A unitary operator $\UU$ that achieves a critical value of $\ep(E_q,U)$ must fulfill
\begin{equation}
\label{Eq.Crit.1st.der}
    \Tr \left( \UU \NN \UU^{\dagger} \left[ 
    \MM_q , \mathcal{G}_a
    \right] \right) = 0 , \quad \mathcal{G}_a = \mathds{1} \otimes G_a + G_a \otimes \mathds{1} ,
\end{equation}
for any generator of the Lie algebra $G_a \in \mathfrak{su}(2j+1)$. If it satisfies additionally that the Hessian of $\ep$ evaluated at $\UU$,
\begin{equation}
\label{Eq.Crit.2nd.der}
    H_{ab} \equiv - \Tr \left( \UU \NN \UU^{\dagger} \left[ \left[ 
    \MM_q , \mathcal{G}_a
    \right] , \mathcal{G}_b
    \right] \right) ,
\end{equation}
has only negative eigenvalues, then $\UU$ is a local maximum. Due to the invariance of $\ep(E_q ,U)$ under left and right SU$(2)$ operations~\eqref{Eq.invariance.ep}, at least 6 eigenvalues of $H$ are equal to zero.
\begin{table}[t]
    \centering
    \begin{tabular}{c | @{\hskip 0.1in}c@{\hskip 0.1in} | c}
      $j$ & $q$
      & $\overrightarrow{\MM}_q $ 
      \\[7pt]
      \hline      
       1  & 1
        & $\left( -2,-\sqrt{3},\sqrt{5} \right)$
       \\[7pt]
       3/2 & 1
       & $\left( -\frac{5}{3},-\frac{11}{3 \sqrt{3}},-\frac{\sqrt{5}}{3},\sqrt{7} \right)$
       \\[7pt]
       2 & 1
       & $
       \left( -\frac{3}{2},-\frac{5 \sqrt{3}}{4} ,-\frac{3 \sqrt{5}}{4} ,0,3 \right)
        $
       \\[7pt]
       & 2 & $\left( -\frac{1}{4},-\frac{\sqrt{3}}{2},-\frac{3\sqrt{5}}{4} ,-\frac{\sqrt{7}}{2},3 \right)$
       \\[7pt]
       5/2  & 1
        &
        $ \left( -\frac{7}{5},-\frac{31 \sqrt{3}}{25} ,-\frac{23}{5 \sqrt{5}},-\frac{11 \sqrt{7}}{25} ,\frac{3}{5},\sqrt{11} \right) $
       \\[7pt]
         & 2
       & $\left( -\frac{7}{20},-\frac{47 \sqrt{3}}{100} ,-\frac{31}{10 \sqrt{5}},-\frac{31 \sqrt{7}}{50} ,-\frac{3}{5},\sqrt{11} \right) $
    \end{tabular}
    \caption{Vectors of SU(2) invariants of $\MM_q$~\eqref{Eq.Exp.Mq.PL} for the spin values $j=1 ,3/2 , 2,5/2$ and $q=1, \dots , \lfloor j \rfloor$. }
    \label{Table.ep}
\end{table}
\section{\texorpdfstring{$\ep$}{Lg} for small number of qubits}
\label{Sec.Examples}
\subsection{ \texorpdfstring{$N=2$ ($j=1$)}{Lg}}
\begin{figure}[t]
\includegraphics[angle=0,width=.4\textwidth]%
{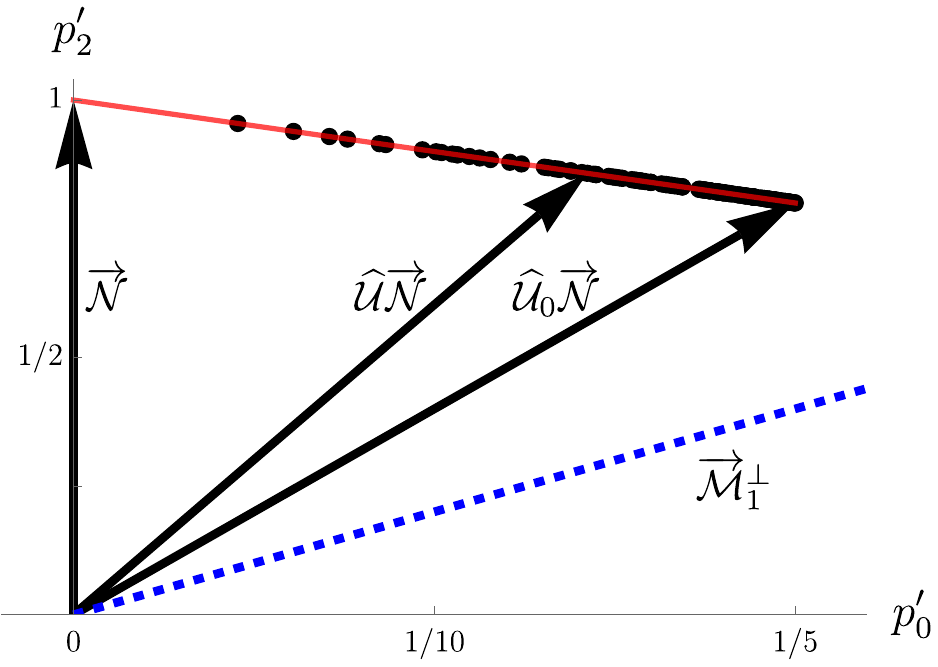}%

\includegraphics[angle=0,width=.4\textwidth]%
{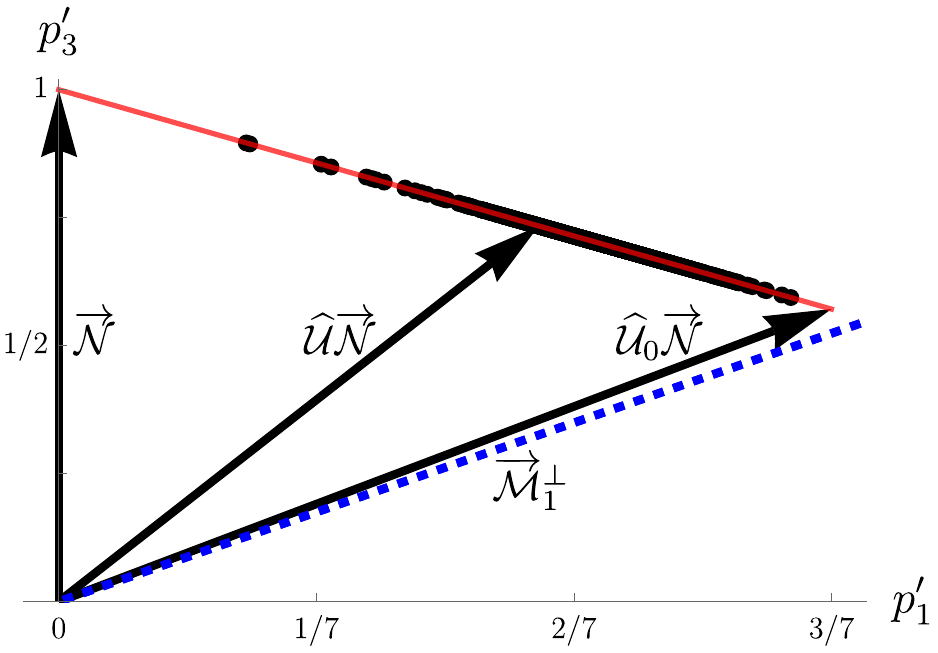}%
\caption{ Vectors of SU(2) invariants in the plane where $\MUU \overrightarrow{\NN}$ has non-zero components~\eqref{Eq.comp.P} for $j=1$ (upper frame) and $j=3/2$ (lower frame). The black dots are the vectors corresponding to random, according to the Haar measure, unitary operators. The solid red segment is the part of the hyperplane~\eqref{Eq.Hyperplane.P} that also satisfies~\eqref{Eq.bounds.components.P2j}. The contour lines of $\ep(E_1,U)$ are parallel to the dashed blue line representing $\overrightarrow{\MM}_1^{\perp}$. The entangling power of $U$ increases as the euclidean distance from $\widehat{U}\overrightarrow{\NN}$ to $\overrightarrow{\MM}_1^{\perp}$ decreases. The upper bounds are saturated for the unitary transformations~\eqref{U0M} (also \eqref{Eq.j1.U0p}) for $j=1$ and~\eqref{Eq.s3h2.U0}, for $j=3/2$. To simplify the  axes labels, we use the scaled variables $p_{\si}'= \sqrt{2\si+1}p_{\si}$.
}
\label{Fig.Ent.pow.spin.1.3h2}
\end{figure}

We calculate again $\ep(E_1 ,U)$ for $j=1$ using the formulation given in the previous section. In this case, $\MUU \overrightarrow{\NN} = (p_0 , 0, p_2)$. The restriction to the hyperplane~\eqref{Eq.Hyperplane.P}, $p_0+\sqrt{5}p_2 =1$, lets us write $\ep$ in terms of only $p_0$,
\begin{equation}
\label{ep.spin1}
  \ep(E_1 ,U) = 1- \left\langle \MUU \overrightarrow{\NN} , \overrightarrow{\MM}_1 \right\rangle = 3 p_0
\end{equation}
We recover~\eqref{Eq.ep.spin1} by calculating $p_0$,
\begin{equation}
\label{I0.inv}
    \begin{aligned}
   5 p_0 = & \Tr \left(  \UU \PP_2 \UU^{\dagger} \PP_0 \right)
    \\
    = & \Tr \left(  \UU (\mathds{1}-\PP_0 - \PP_1) \UU^{\dagger} \PP_0 \right)
\\
= & 1- \Tr \left(  \UU \PP_0  \UU^{\dagger} \PP_0 \right) .
    \end{aligned}
\end{equation}
Since $\PP_0$ is a rank one operator, the last expression can be rewritten as
\begin{equation}
   5 p_0 = 
1 - \left|\Tr \left( \PP_0 \UU \right) \right|^2
=
1 - \frac{1}{9} \left| \Tr \left( \Phi U^T \Phi U \right) \right|^2
    ,
\end{equation}
where $\Phi$ is defined in Eq.~\eqref{m.exp1} and the last equality is derived as follows
\begin{equation}
    \begin{aligned}
\Tr (\PP_0 \UU) = & \sum_{ \substack{        m_1 \, m_2     \\
        n_1 \, n_2   
}}
    C_{1 m_1 1 m_2}^{00}
    C_{1 n_1 1 n_2}^{00} U_{m_1 n_1} U_{m_2 n_2} 
    \\
    = & \, \frac{1}{3}
    \sum_{ m_1 \, n_1 } 
    (-1)^{m_1 + n_1}
     U_{m_1 n_1} U_{-m_1 -n_1}  
\\
    = & \, \frac{1}{3} 
    \Tr \big( \Phi U^T \Phi U \big)
     .
    \end{aligned}
\end{equation}
We can obtain the upper bound $\ep(E_1,U) \leqslant 3/5 $ from Eq.~\eqref{Eq.bounds.components.P2j} --- the inequality is in fact saturated by $U_0$~\eqref{U0M}. 
In Fig.~\ref{Fig.Ent.pow.spin.1.3h2} (upper frame) we plot the vectors $\MUU \overrightarrow{\NN}$ for random unitary operators, produced with the Haar measure, as well as the vector $\MUU_0 \overrightarrow{\NN}$ with $U_0$ given by~\eqref{U0M}. We also plot the orthogonal complement $\overrightarrow{M}_q^{\perp}$ of $\overrightarrow{M}_q$ with respect to the inner product~\eqref{Eq.Inner.Product}. As expected, $\UU_0$ meets the criteria for a local maximum~\eqref{Eq.Crit.1st.der}-\eqref{Eq.Crit.2nd.der}, with the Hessian there having two eigenvalues equal to $-4$ and six equal to $0$. 
%
%
%%%%%%%%%%%%%%%%%% BEGIN FIGURE
%%%%%%%% END FIGURE
%
\subsection{\texorpdfstring{$N=3$ ($j=3/2$)}{Lg}}
The vector $\MUU \overrightarrow{\NN} = (0,p_1,0,p_3)$ is restricted in the hyperplane~\eqref{Eq.Hyperplane.P}
$ \sqrt{3}p_1 +  \sqrt{7}p_3 = 1  $.
Hence, $\ep(E_1 ,U)$ is also a function of one SU$(2)$ invariant for $j=3/2$, 
\begin{equation}
\label{ep.spin3h2}
\begin{aligned}
    \ep(E_1 ,U)
    = 
    \frac{20\sqrt{3}}{9} p_1 
    \leqslant
    \frac{20}{21}
    .
\end{aligned}
\end{equation}
The bound is saturated by the unitary operator
\begin{equation}
\label{Eq.s3h2.U0}
U_0 = \left(
\begin{array}{cccc}
 0 & 1 & 0 & 0 \\
 0 & 0 & 0 & i \\
 i & 0 & 0 & 0 \\
 0 & 0 & 1 & 0 
\end{array}
\right) ,
\end{equation}
which we identified as outlined in Section~\ref{Sec.Schmidt}. Note that, in order to simplify the notation, we denote the optimal entangles by $U_0$ for all values of spin --- which unitary operator is involved should be clear from the context. We plot the vectors $\MUU \overrightarrow{\NN}$ for random unitary operators, as well as $\MUU_0 \overrightarrow{\NN}$ and $\overrightarrow{\MM}_q^{\perp}$ in Fig.~\ref{Fig.Ent.pow.spin.1.3h2} (lower frame). Again, it is verified that $U_0$ is a critical value of $\ep$ with Hessian having eigenvalues equal to $-8$, $-8/5$ and 0, with degeneracies 4, 5 and 6, respectively. 
\subsection{\texorpdfstring{$N=4$ ($j=2$)}{Lg}}
\begin{figure}[t]
\includegraphics[angle=0,width=.45\textwidth]%
{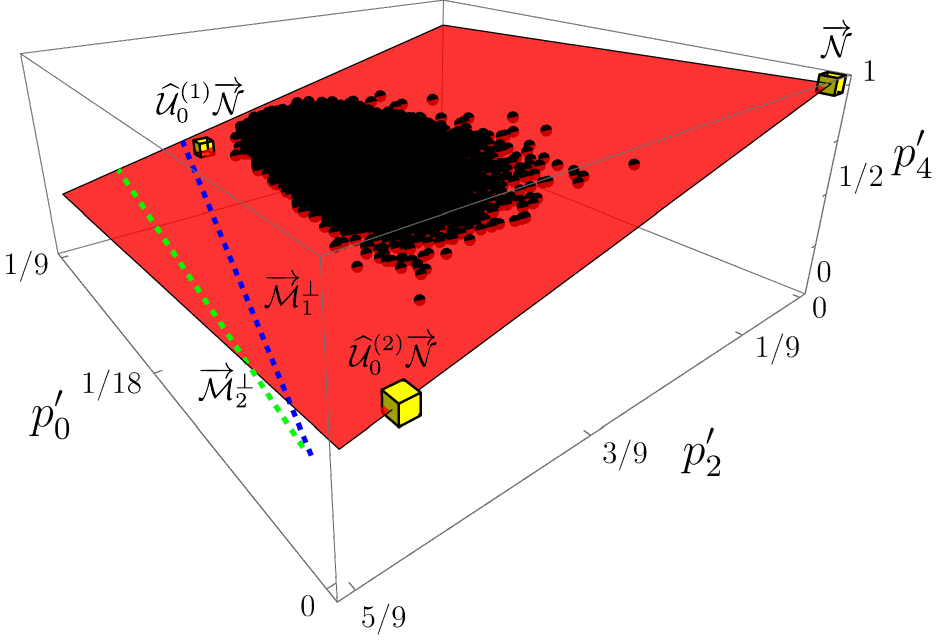}%
\caption{ Vectors $\MUU \overrightarrow{\NN}$ for $j=2$, where we plot only their non-zero components~\eqref{Eq.comp.P}. The black dots are the vectors given by random unitary operators. The red plane is the restriction~\eqref{Eq.Hyperplane.P}. The dashed blue and green lines denotes the intersection between the red plane with $\overrightarrow{\MM}_1^{\perp}$ and $\overrightarrow{\MM}_2^{\perp}$, respectively. The yellow cubes represent the position of the vector $\overrightarrow{\NN}$ and its corresponding transformed vector by the unitary operators $U_0^{(q)}$ given by Eqs.~\eqref{Eq.U0q1j2} and \eqref{Eq.U0q2j2}, respectively.
}
\label{Fig.EP.N4}
\end{figure}

Here, we have three different non-zero components in $\MUU \overrightarrow{N} = \left( p_0 ,0,p_2 ,0 ,p_4 \right)$ restricted to the plane $p_0+\sqrt{5}p_2+3 p_4=1$, and two different non-equivalent bipartite entanglements
\begin{equation}
\label{Eq.EP.N4}
\begin{aligned}
    \ep (E_q,U)= & \frac{1}{4} \left( \frac{10p_0}{q}+7\sqrt{5}p_2 \right) ,
\end{aligned}
\end{equation}
for $q=1,2$. Unlike the previous cases, the inequalities~\eqref{Eq.bounds.components.P2j} provide a trivial bound for $\ep(E_q,U)$. Through numerical search, we identified unitary matrices $U_0^{(q)}$ that we conjecture are optimal entanglers for  $E_q$, $q=1,2$. They read
\begin{equation}
\label{Eq.U0q1j2}
U_0^{(1)}=
\left(
\begin{array}{ccccc}
   \beta  \cos \alpha  & 0 & 0 & 0 & i  \beta \sin \alpha  \\
 0 & e^{i\frac{\pi}{4} } & 0 & 0 & 0 \\
 0 & 0 & \beta^2 & 0 & 0 \\
 0 & 0 & 0 & e^{-i\frac{\pi}{4} } & 0 \\
-i  \beta  \sin \alpha & 0 & 0 & 0 & - \beta  \cos \alpha  \\
\end{array}
\right) ,
\end{equation}
with $\alpha= \arctan \left(\sqrt{83/53}\right)$ and $\beta= e^{-i \arctan\left(\sqrt{53/83}\right)}$, and
\begin{equation}
\label{Eq.U0q2j2}
    U_0^{(2)} = \left(
\begin{array}{ccccc}
 0 & 0 & 0 & i & 0 \\
 i & 0 & 0 & 0 & 0 \\
 0 & 0 & 1 & 0 & 0 \\
 0 & 0 & 0 & 0 & i \\
 0 & i & 0 & 0 & 0 \\
\end{array}
\right) .
\end{equation}
Both unitary matrices fulfill the criteria for local maximum~\eqref{Eq.Crit.1st.der} and \eqref{Eq.Crit.2nd.der} for their respective $E_q$, with entangling power 
\begin{equation}
\begin{aligned}    
    \ep \left( E_1,U_0^{(1)} \right) = & \frac{6889}{7140} \approx 0.9648 ,
    \\
    \ep \left( E_2,U_0^{(2)} \right) = & \frac{25}{28} \approx 0.8929 .
\end{aligned}
\end{equation}
As for the previous cases, we plot the vectors $\MUU \overrightarrow{\NN}$, $\MUU_0^{(q)} \overrightarrow{\NN}$ and $\overrightarrow{\MM}_q^{\perp}$ in Fig.~\ref{Fig.EP.N4}.
\subsection{\texorpdfstring{$\ep$}{Lg} of a unitary operator and its inverse}
In this short section we comment on the relation between the entangling power of a unitary matrix $U$ and its inverse. We find that, for $j=1$ and $j=3/2$, $\ep(E_1 , U) = \ep(E_1 , U^\dagger )$ for all  $U \in \mathrm{SU}(2j+1)$. These results are derived from the expressions 
\begin{equation}
    \begin{aligned}
\ep (E_1 , U) =  \frac{3}{5} \left[1- \Tr \left(\UU \PP_0 \UU^{\dagger} \PP_0 \right) \right] ,
\quad & \text{ for } j=1,
        \\
\ep (E_1 , U) =  \frac{20}{63} \left[3- \Tr \left(\UU \PP_1 \UU^{\dagger} \PP_1 \right) \right] ,
\quad & \text{ for } j=3/2,    \end{aligned}
\end{equation}
obtained from Eqs.~\eqref{ep.spin1}-\eqref{I0.inv} and Eq.~\eqref{ep.spin3h2}, respectively. On the other hand, numerical calculations show that, in general, $\ep(E_q , U) \neq \ep(E_q , U^\dagger )$ for $j\geqslant 2$. In the case of $j=2$, for instance, the $\ep$~\eqref{Eq.EP.N4} can be rewritten as
\begin{equation}
\begin{aligned}
&    36\ep(E_q, U)
=  
- \frac{10}{q} \Tr \left( \UU \PP_2 \UU^{\dagger} \PP_0 \right) - 7 \Tr \left( \UU \PP_0 \UU^{\dagger} \PP_2 \right)  
 \\
&+ \frac{10}{q}\Big(1 -  \Tr \left( \UU \PP_0 \UU^{\dagger} \PP_0 \right) \Big)  
 + 7 \Big(5 -  \Tr \left( \UU \PP_2 \UU^{\dagger} \PP_2 \right) \Big) 
\end{aligned}
\end{equation}
where the first two terms on the right-hand side provide the difference between $\ep(E_q , U)$ and $\ep(E_q , U^\dagger )$.
%
%%%%%%%%%%%%%%%%%% BEGIN FIGURE
%
%
%
%
%%%%%%%%%%%%%%%%%END FIGURE
%
\begin{figure*}[t]
\begin{center}    
a) \hspace{5.2cm} b) \hspace{5.2cm} c)
\\[0.2cm]
\mbox{
\includegraphics[angle=0,height=.24\textwidth]{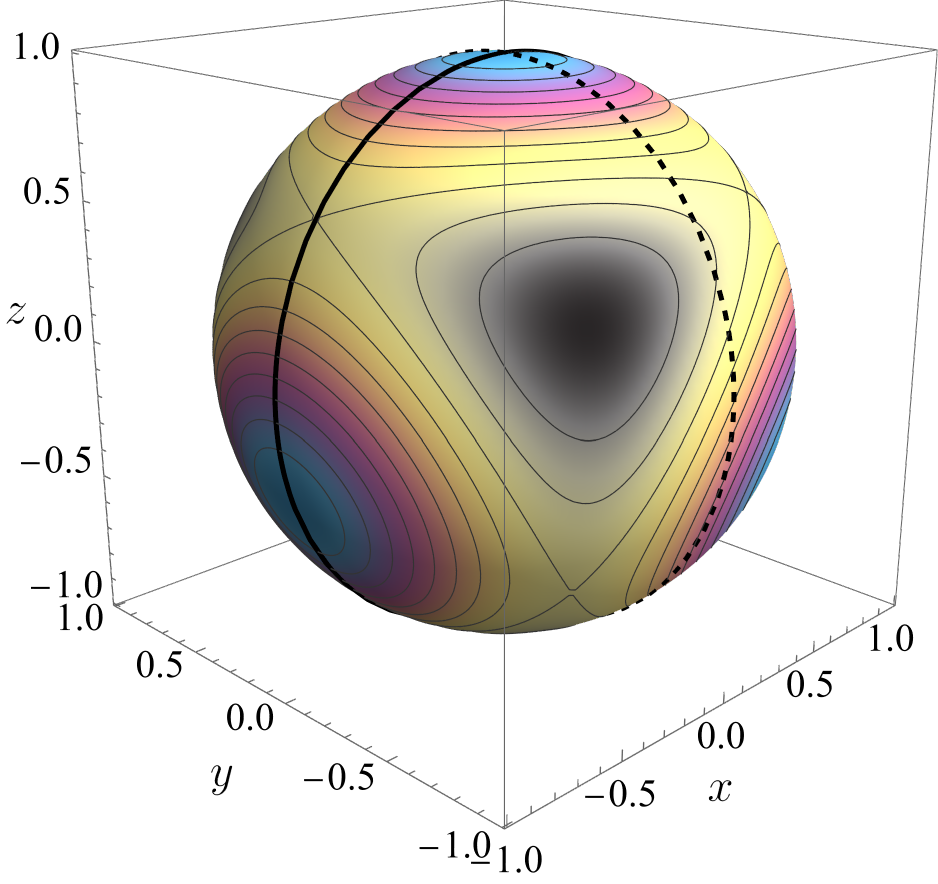}%
\includegraphics[angle=0,height=.24\textwidth]{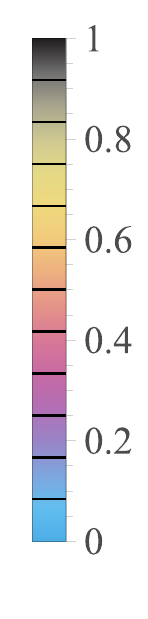}%
\hspace{0.05cm}
\includegraphics[angle=0,height=.24\textwidth]{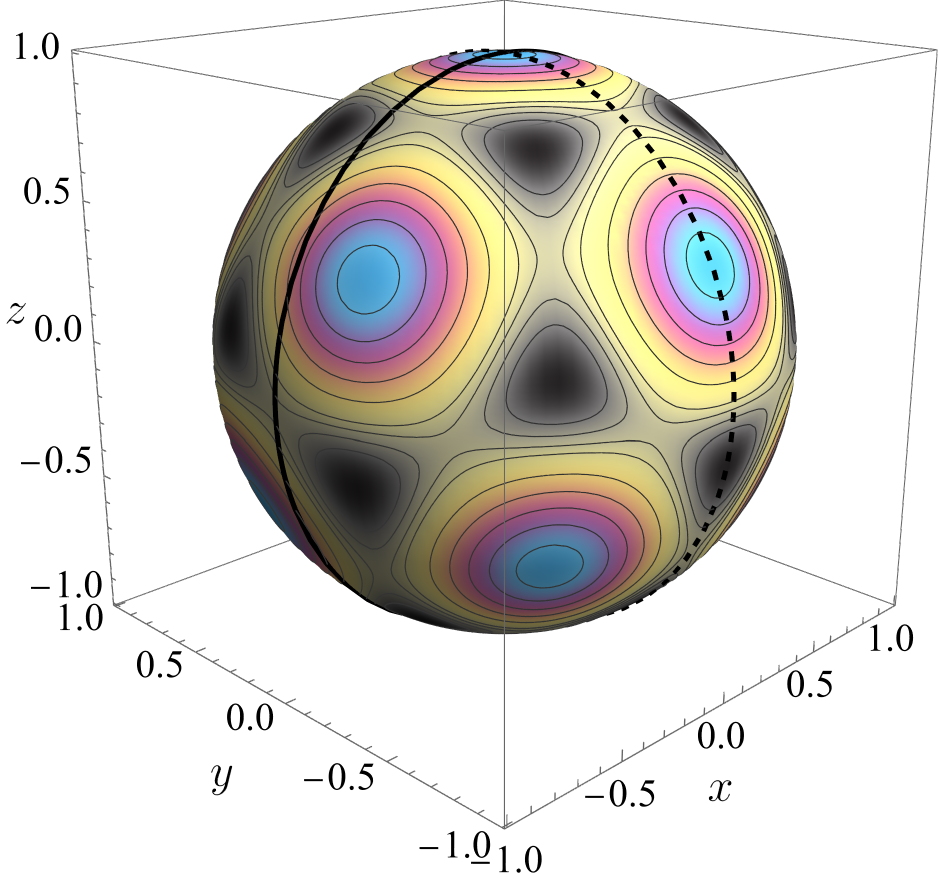}
\includegraphics[angle=0,height=.24\textwidth]{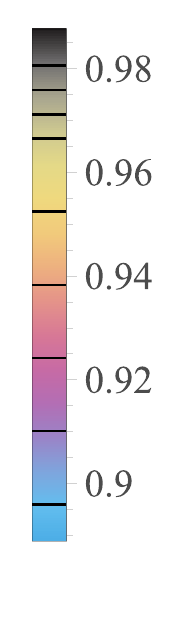}%
\hspace{0.05cm}
\includegraphics[angle=0,height=.24\textwidth]{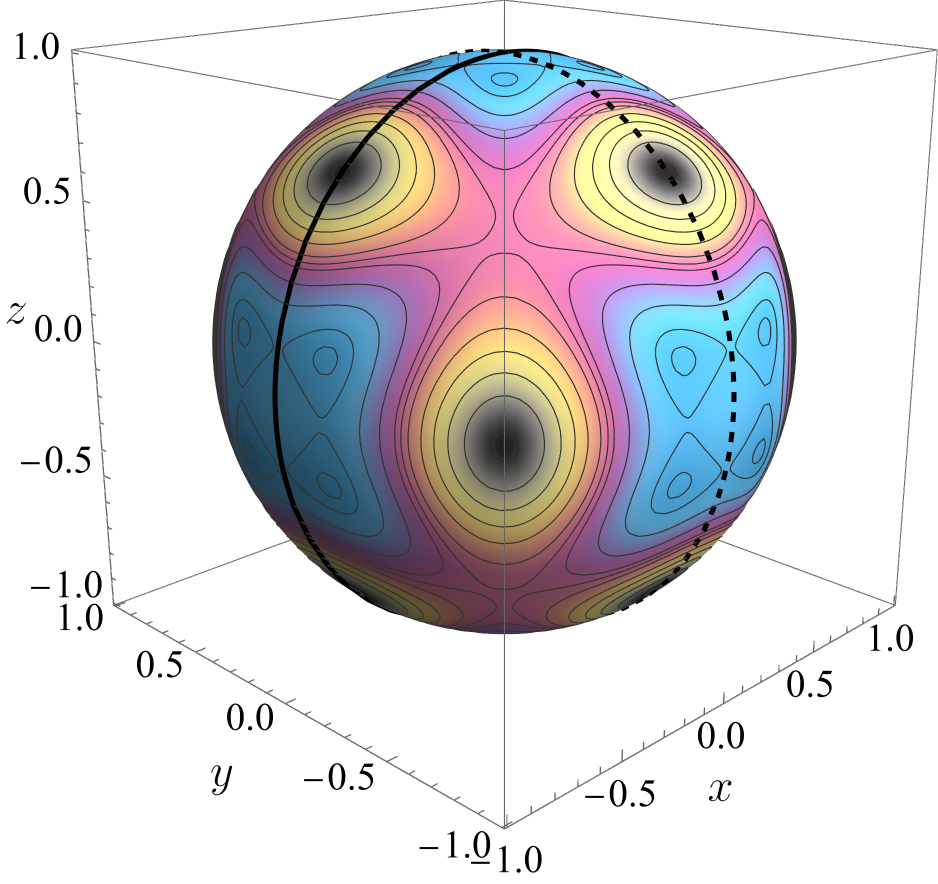}%
\includegraphics[angle=0,height=.24\textwidth]{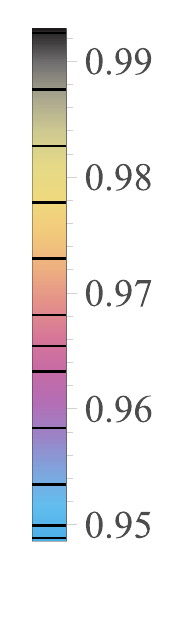}%
}
\\
\includegraphics[angle=0,height=.32\textwidth]{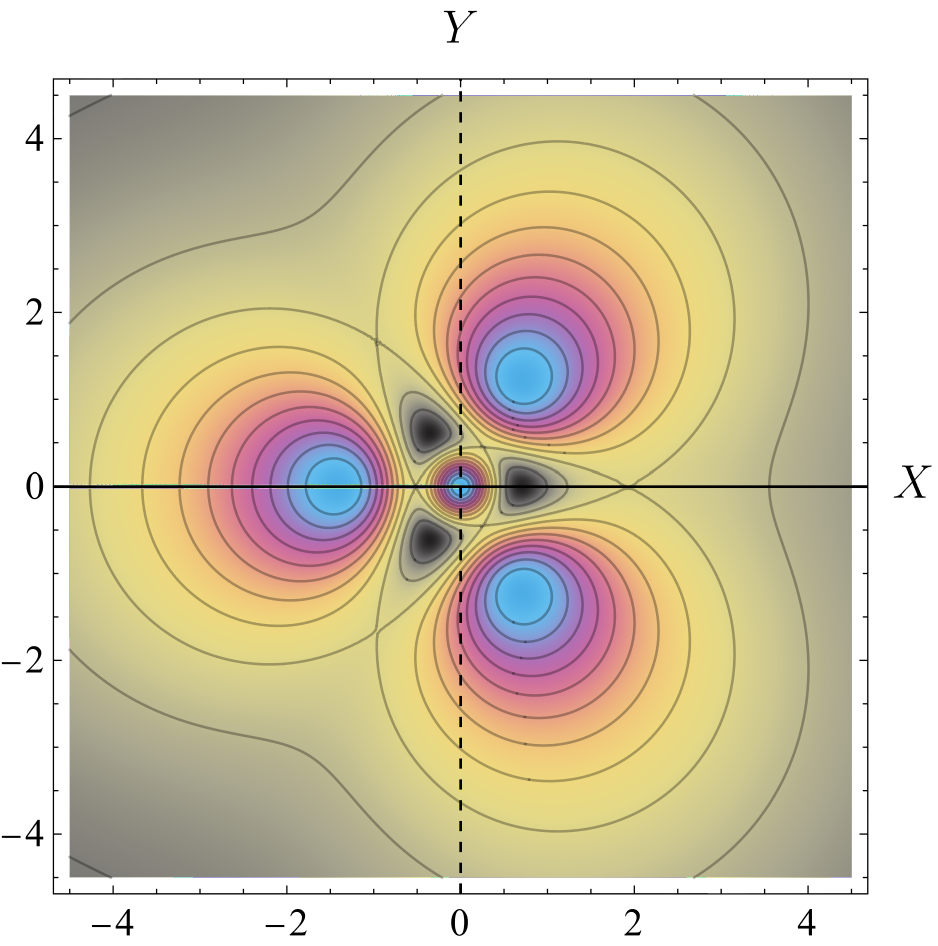}%
\hspace{0.1cm}
\includegraphics[angle=0,height=.32\textwidth]{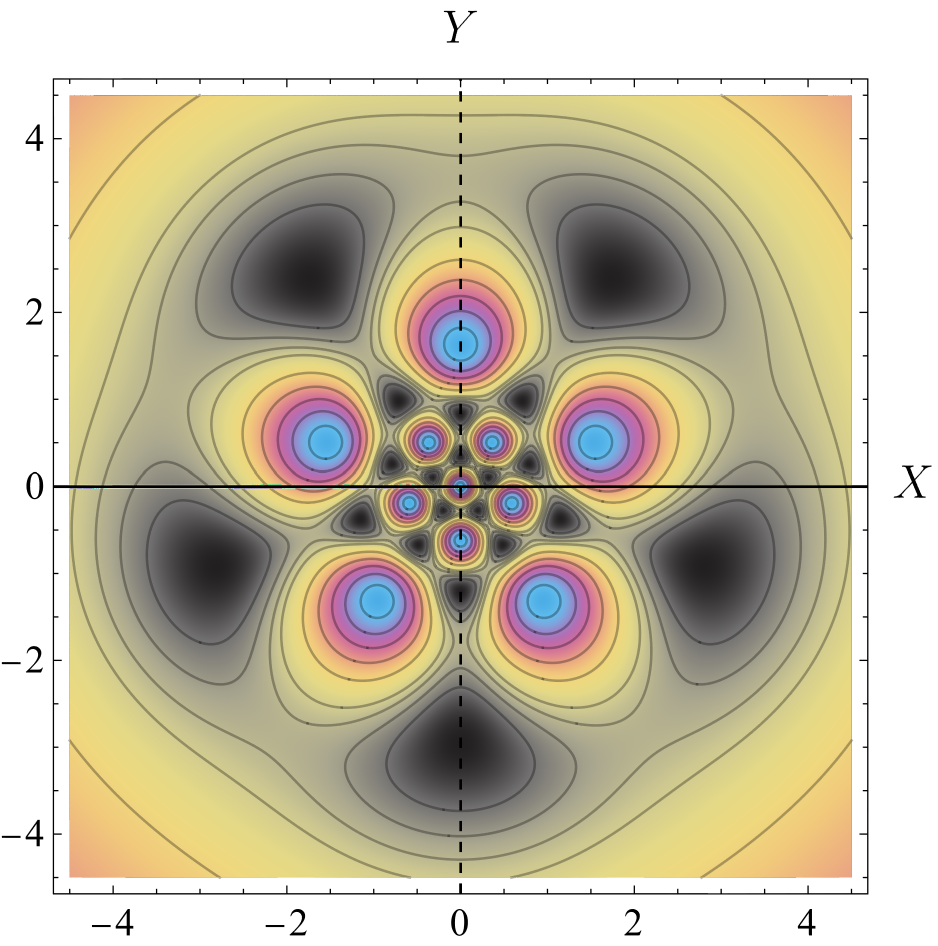}%
\hspace{0.1cm}
\includegraphics[angle=0,height=.32\textwidth]{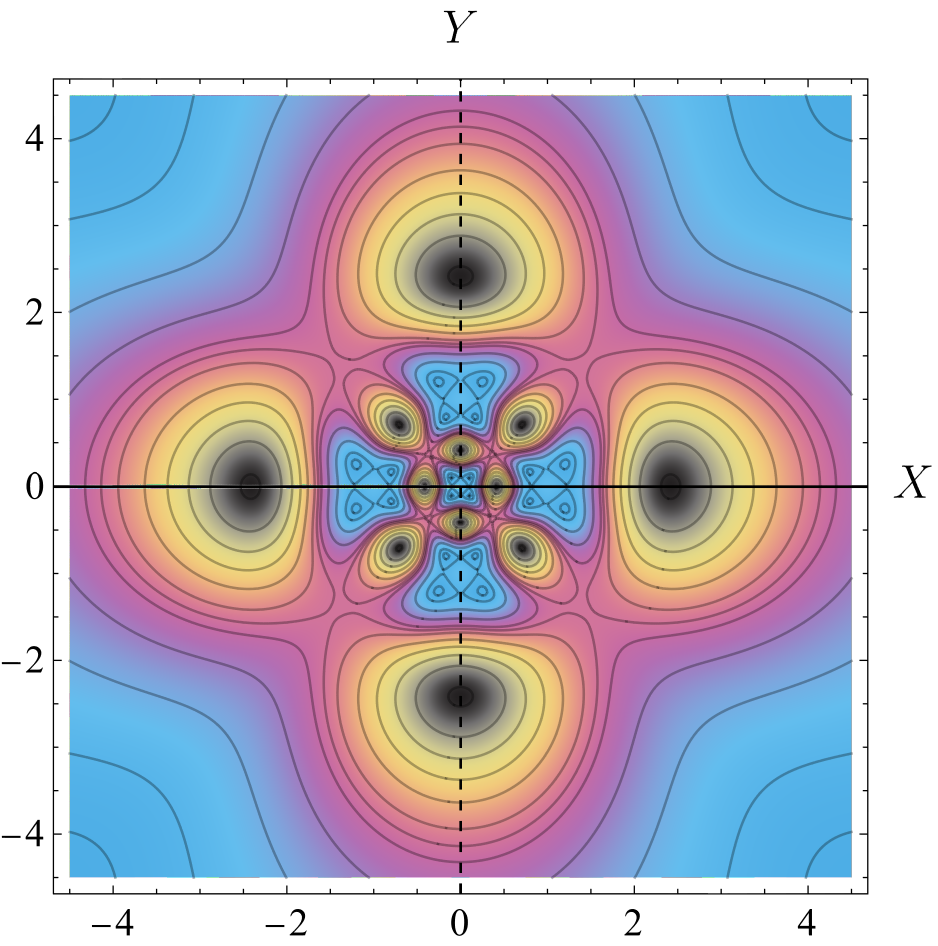}%
\end{center}
\caption{Entanglement distribution $E_1(U\ket{ j,\mathbf{n}})$ (Top) and its stereographic projection (Bottom) obtained by the SC states after the action of the unitary operator a) \eqref{Eq.j1.U0p} b) \eqref{Eq.s3h2.U0} and c)~\eqref{Eq.U0q1j2} for the spin values $j=1,3/2, 2$, respectively. $E_q(U\ket{ \mathbf{n}})$ has tetrahedral, icosahedral and octahedral symmetry, respectively. These unitary operators maximize $\ep$ for $j=1,3/2$, and apparently for $j=2$, respectively. The numerical values of the contour lines are marked in each color bar.}
\label{fig.Ent.Dis.Tod}
\end{figure*}
%
%%%%%%%%%%%%%%%%%BEGIN FIGURE
%
\section{Entanglement distribution and Husimi functions}
\label{Sec.Husimi}
Now, we would like to study the entanglement $E_q(U\ket{j,\mathbf{n}})$ as a function on the sphere where $\mathbf{n}$ lives.
We start with the expression
\begin{equation}
\label{Eq.ED}    
E_q(U\ket{j,\mathbf{n}}) =1- \bra{2j,\mathbf{n}} \ED_q (U)   \ket{2j,\mathbf{n}} ,
\end{equation}
with
\begin{equation}
  \ED_q (U)  \equiv \PP_{2j} \UU^{\dagger} \MM_{q} \UU \PP_{2j} \, ,
\end{equation}
which follows from~\eqref{Eq.Ent.q.first}
since the state $\ket{2j,\mathbf{n}}$ lies in the subspace associated to $\PP_{2j}$. The matrix $\ED_q = \ED_q (U)$ is Hermitian for any $U$ and can be thought of as a $(4j+1)\times (4j+1)$ matrix when restricted to the image of $\PP_{2j}$. Additionally, it rotates under $\mathrm{SU}(2)$ transformations as $D^{(2j)}(R) \ED_q D^{(2j) \dagger}(R)$ with $D^{(2j)}(R)$ the $(2j)$-irrep of the rotation $R$. By Eq.~\eqref{Eq.ED}, $E_q(U \ket{j,\mathbf{n}})$ inherits the rotational symmetries of $\ED_q$. These symemtries can be scrutinized by the Majorana representation for Hermitian operators~\cite{Ser.Bra:20}. In particular, we use it to verify the rotational symmetries of $E_q(U \ket{j,\mathbf{n}})$ for the unitary gates mentioned below. 
 
The eigendecomposition  $\ED_q = \sum_{k=1}^{4j+1} \si_{k} \ket{\psi_k} \bra{\psi_k}$ also helps to recast the entanglement distribution on the sphere as
\begin{equation}
    E_q (U  \ket{j,\bm{n}}) = 1 - \sum_{k=1}^{4j+1} \si_k H_{\ket{\psi_k}} (\mathbf{n}) ,
\end{equation}
with $H_{\ket{\psi_k}} (\mathbf{n}) =  |\langle 2j, \mathbf{n} | \psi_k \rangle|^2$ being the Husimi function of $\ket{ \psi_k}$~\cite{PhysRevA.24.2889}. By averaging over the sphere, we find
\begin{equation}
 \begin{aligned}     
    \ep (E_q , U) = & 1- \sum_{k=1}^{4j+1} \si_k \overline{H_{\ket{\psi_k}} (\mathbf{n})} 
    \\
    = &
    1- \frac{1}{4j+1} \sum_{k=1}^{4j+1} \si_k 
    \\
    = &
    1- \frac{\Tr \left[ \ED_q (U) \right]}{4j+1} 
    ,
 \end{aligned}
\end{equation}
which reduces to Eq.~\eqref{Eq.EP.vec}. Let us now study the entanglement distribution of the examples in Sec.~\ref{Sec.Examples}.
\subsection{\texorpdfstring{$N=2$ ($j=1$)}{Lg}}
We plot in Fig.~\ref{fig.Ent.Dis.Tod}a the entanglement distribution
\begin{equation}
\label{En1U}
\begin{aligned}
& E_1(U_0'\ket{ \mathbf{n}})=
4 \sqrt{2} \sin ^5\left(\frac{\theta }{2}\right) \cos ^3\left(\frac{\theta }{2}\right) \cos (3 \phi )
\\
& + \frac{\sin ^2\left(\frac{\theta }{2}\right)}{32} \big[ 90+ 105 \cos (\theta )+54 \cos (2 \theta )+7 \cos (3 \theta ) \big]
\, ,
\end{aligned}
\end{equation}
with $U_0'$ defined in~\eqref{Eq.j1.U0p}.
Its corresponding matrix $\ED_1$ is equal to
\begin{equation}
    \ED_1 = \left(
\begin{array}{ccccc}
 1 & 0 & 0 & 0 & 0 \\
 0 & -1 & 0 & 0 & \sqrt{2} \\
 0 & 0 & 1 & 0 & 0 \\
 0 & 0 & 0 & 1 & 0 \\
 0 & \sqrt{2} & 0 & 0 & 0 \\
\end{array}
\right) ,
\end{equation}
which exhibits tetrahedral symmetry. The same point group can be observed in the entanglement distribution of $U_0'$ plotted in Fig.~\ref{fig.Ent.Dis.Tod}a. $U_0'$ does not create any entanglement when applied to four SC states pointing in the vertices of a regular tetrahedron, one of which is $\ket{ \mathbf{z}}$. On the other hand, it transforms the SC states pointing along the vertices of the antipodal tetrahedron into maximally entangled states. For instance, $U_0' \ket{-\mathbf{z}} = \ket{1,0}$. An alternative way to show the tetrahedral symmetry of the entanglement distribution is by direct calculation of the eigendecomposition of $\ED_1$
\begin{equation}
    \ED_1 = \mathds{1}_5 -3 \ket{\psi_{\text{T}}} \bra{\psi_{\text{T}}}  
\end{equation}
with
\begin{equation}
\ket{\psi_T} = - \sqrt{\frac{2}{3}} \ket{2,1}  + \frac{1}{\sqrt{3}} \ket{2,-2}  
\end{equation}
a spin-2 state with tetrahedral symmetry~\cite{Bag.Mar:17}. By direct algebra, we obtain that 
\begin{equation}
    E_1 (U_0', \ket{1,\mathbf{n}}) =  3 H_{\ket{\psi_T}} ( \mathbf{n}) ,
\end{equation}
\ie, the entanglement distribution of $U_0'$ is proportional to the Husimi function of the tetrahedron state $\ket{\psi_T}$.

Similar expressions for $\ED_1$ and $E_1$ are obtained for a general $U \in \mathrm{SU}(3)$ using the parametrization~\eqref{Ec:Anonlocal}. By taking only the non-local term of the unitary gate $U=A$, we get that
\begin{equation}
    \ED_1 = \mathds{1}_5 - \frac{4\Gamma_A}{3}  \ket{\psi_A} \bra{\psi_A} ,
\end{equation}
with 
\begin{equation}
    \Gamma_A= \sin^2 c_{12} + \sin^2 c_{13} + \sin^2 c_{23} 
\end{equation}
and
\begin{equation}
\label{Eq.State.A}
\begin{aligned}    
    \ket{\psi_A} =&  
    \sqrt{\frac{3}{\Gamma_A}} \frac{\sin c_{12}}{2} \Big( \ket{2,2} + \ket{2,-2}  \Big)
    \\
   & +i \left( \frac{\cos c_{12} - e^{-i(c_{13}+c_{23})}}{\sqrt{2\Gamma_A}}
    \right) \ket{2,0} .
\end{aligned}
\end{equation}
Thus, 
\begin{equation}
\label{Eq.j1.ED}
    E_1 (U, \ket{1,\mathbf{n}}) =  \frac{4\Gamma_A}{3} H_{\ket{\psi_A}} ( \mathbf{n}) ,
\end{equation}
with $\Gamma_A$ proportional to $\ep(E_1, U)$ \eqref{epc1c2}~\footnote{
Using the multipole expansion~\eqref{rhoTLMexpansion}, each spin-1 operator has associated to it a spin-2 state. Specifically, the components $\bm{\rho}_2 (U)$ can be expressed as $\ket{\bm{\rho}_2 (U)} = \sum_{m=-2}^2 \rho_{2m}(U) \ket{2,m}$. In particular, for $U=A \in \mathrm{SU}(3)$, where $A$ is defined in Eq.~\eqref{Ec:Anonlocal}, we obtain that $\ket{\bm{\rho}_2 (A^{*2})} \propto \ket{\psi_{A}}$ (see Eq.~\eqref{Eq.State.A}). However, this proportionality does not extend to higher spins.}.
\subsection{\texorpdfstring{$N=3$ ($j=3/2$)}{Lg}}
We now plot $E_1(U_0\ket{3/2,\mathbf{n}})$, with $U_0$ the unitary transformation in~\eqref{Eq.s3h2.U0}, in Fig.~\ref{fig.Ent.Dis.Tod}b. The entanglement distribution  
\begin{equation}
\label{Eq.Ent.D.s3h2}
\begin{aligned}    
    E_1(U_0 \ket{3/2,\mathbf{n}}) = & \frac{1}{1152} \left[ 1090+
192 \sin ^5(\theta ) \cos (\theta ) \sin (5 \phi ) \right.
\\ & \left.
-15 \cos (2 \theta ) -18 \cos (4 \theta )-33 \cos (6 \theta ) \right]
    \end{aligned}
\end{equation}
has an icosahedral symmetry, and
takes values in the interval $[8/9 , 80/81]$.
The rotational symmetries are also reflected in the configuration of the minima and maxima, which are arranged in an icosahedron and a dodecahedron, respectively. 
The corresponding $\ED_1$ matrix has eigendecomposition given by
\begin{equation}
    \ED_1 = \mathds{1}_7 -\frac{20}{9} \sum_{k=1}^3 \ket{\psi_k} \bra{\psi_k} ,
\end{equation}
with states~\footnote{The spin-3 states shown in Eq.~\eqref{Eq.Bases.N3} span a 3-dimensional 1-anticoherent subspace (see Ref.~\cite{Ser.Chr.Mar:24} for more details).}
\begin{equation}
\label{Eq.Bases.N3}
\begin{aligned}
    \ket{\psi_1} = & - i \sqrt{\frac{3}{5}} \ket{3,2} + \sqrt{\frac{2}{5}} \ket{3,-3} ,
\\    
    \ket{\psi_2} = & i \sqrt{\frac{2}{5}} \ket{3,3} + \sqrt{\frac{3}{5}} \ket{3,-2} ,
\\
    \ket{\psi_3} = & \ket{3,0} .
\end{aligned}
\end{equation}
The entanglement distribution is then given by
\begin{equation}
     E_1(U_0 \ket{3/2,\mathbf{n}})  = \frac{20}{9} \sum_{k=1}^3 H_{\ket{\psi_k}} (\mathbf{n}) .
\end{equation}
\subsection{\texorpdfstring{$N=4$ ($j=2$)}{Lg}}
We plot $E_1(U_0^{(1)}, \ket{2,\mathbf{n}})$, with $U_0^{(1)}$ defined by~\eqref{Eq.U0q1j2}, in Fig.~\ref{fig.Ent.Dis.Tod}c. We observe octahedral symmetry, confirmed by using the Majorana representation of mixed states~\cite{Ser.Bra:20}. 
In fact, the minima (maxima) of $E_1(U_0^{(1)} \ket{2,\mathbf{n}})$ lie on the vertices of a truncated octahedron (cuboctahedron) --- the explicit expression of this function is rather long and is not particularly enlightening. On the other hand, $E_2(U_0^{(2)}  \ket{2,\mathbf{n}})$, with  $U_0^{(2)}$ as in Eq.~\eqref{Eq.U0q2j2}, is given by an affine transformation of the entanglement distribution plotted in Fig.~\ref{fig.Ent.Dis.Tod}b,
\begin{equation}
    4 E_2(U_0^{(2)} \ket{2,\mathbf{n}}) =  9 E_1 (U_0^{\text{Eq.}\eqref{Eq.s3h2.U0}}  \ket{3/2,\mathbf{n}}) -5
    .
\end{equation}
Thus, $E_2(U_0^{(2)} \ket{2,\mathbf{n}})$ has icosahedral symmetry.
\section{Average of \texorpdfstring{$\ep$}{Lg} over the unitary orbit}
\label{Sec.Average}
We calculate the average of $\ep(E_q,U)$ over the unitary operators SU$(d)$, with $d=2j+1$, and with respect to the normalized Haar measure ($\int \diff \mu (U)=1$)
\begin{equation}
\label{Eq.Average.Ep}    
\begin{aligned}    
\overline{\ep(E_q,U) }^{\mathrm{SU}(d)} \equiv  \int \ep(E_q,U) \diff \mu(U) & = 1 -\frac{1}{2j+1-q} ,
\end{aligned}
\end{equation}
the details can be found in Appendix~\ref{App.Average.EP}. The last equation can be written in terms of the dimensions of the initial and the reduced Hilbert spaces, $\Hs^{(q/2)} \otimes \Hs^{(j-q/2)}$ and $\Hs^{(q/2)}$ respectively, 
\begin{equation}
\label{Eq.Average.Ep.dim}
\overline{\ep(E_q,U) }^{\mathrm{SU}(d)}  = 1 - \frac{\dim (\Hs^{(q/2)})}{\dim (\Hs^{(q/2)} \otimes \Hs^{(j-q/2)})} .
\end{equation}
We observe that, similar to the non-symmetric case~\cite{PhysRevA.69.052330}, the average of $\ep$ increases as the dimension of the initial (resp.{} reduced) Hilbert space increases (resp.{} decreases). However, this dependence differs from that of the non-symmetric case. To corroborate this, let us briefly review $\ep$ in the non-symmetric case.

We start with a system of $n$ qudits $\Hs_D^{\otimes n}$. Then, we calculate its average linear entanglement entropy after we trace out $n-q$ constituents, $Q_q$ (see~\cite{PhysRevA.69.052330} for the formal definition), where the average here means over all the possible bipartitions $q|n-q$ over the $n$ constituents. In particular, if the state is symmetric, $Q_q = E_q$. Now, the entangling power with respect to a unitary matrix $\mathbb{U} \in \mathrm{SU}(D^n)$, $\ep(Q_q,\mathbb{U})$, is equal to 
\begin{equation}
\label{Eq.EP.NON}
\begin{aligned}
\ep(Q_q,\mathbb{U}) = & \int Q_q \Big( \mathbb{U} \left(\ket{\psi_1} \otimes \ket{\psi_2} \otimes \dots \otimes \ket{\psi_n} \right) \Big)
\\
& \phantom{\int} \times  \diff \mu(\psi_1) \dots \diff \mu(\psi_n)  ,
\end{aligned}
\end{equation}
where $\int \diff \mu (\psi_k)= 1$. It turns out that its average over the unitary orbit SU$(D^n)$ is equal to~\cite{PhysRevA.69.052330}
\begin{equation}
\label{Eq.Ave.Ep.NS}
\overline{\ep(Q_q,\mathbb{U}) }^{\mathrm{SU}(D^n)}  = 1- \frac{\dim (\Hs_D^{\otimes q}) +1 }{\dim (\Hs_D^{\otimes n}) +1}.
\end{equation}
where $\dim (\Hs_D^{\otimes \alpha}) = D^{\alpha}$. We can observe a difference between the previous equation and Eq.~\eqref{Eq.Average.Ep}. However, a different nature between the average of $e_p$'s in the symmetric and non-symmetric case is expected because we integrated over different sets of product states (see Eqs.~\eqref{Eq.entangling.power} and ~\eqref{Eq.EP.NON}) and over different sets of unitary gates (see Eqs.~\eqref{Eq.Average.Ep} and \eqref{Eq.Ave.Ep.NS}). We tabulate $\overline{\ep(E_q,U) } $ for several values of $j$ and $q$ in Table~\ref{Eq.Tab.Av.Ep}. 
\begin{table}[t]
    \centering
    \begin{tabular}{c|@{\hskip 0.1in}c@{\hskip 0.1in}|@{\hskip 0.1in}c@{\hskip 0.1in} |@{\hskip 0.1in}c@{\hskip 0.1in}}
        $j$ & $q$ & $\overline{\ep(E_q,U) }^{\mathrm{SU}(2j+1)}$ & $\max_U \ep(E_q,U)$
        \\[5pt] \hline
        $1$ & $1$ &  $1/2$ & $3/5$
        \\[5pt]
        $3/2$ & $1$ & $2/3$ & $20/21$
        \\[5pt]
        $2$ & $1$ & $3/4$ & $6889/7140$
        \\[5pt]
        $2$ & $2$ &  $2/3$ & $25/28$
    \end{tabular}
    \caption{Average of $\ep(E_q,U) $ over the set of unitary gates SU$(2j+1)$ for several values of $j$ and $q$. Additionally, we include the maximum possible value of $\ep$ for $j=1$ and 3/2, as well as the conjectured maximum values for $j=2$ obtained from numerical searches.}
\label{Eq.Tab.Av.Ep}
\end{table}
\section{\texorpdfstring{$\ep$}{Lg} and Schmidt numbers }
\label{Sec.Schmidt}
The reformulation of $\ep$ in Eq.~\eqref{Eq.EP.vec} shows that the entangling power increases as the subspace associated with the image of $\NN = \PP_{2j}/(4j+1)$, $\IMA(\NN)$, is transformed to the orthogonal complement of $\MM_q^{\perp}$. The Schmidt numbers of the states $\ket{\psi} \in \Hs^{(j) \otimes 2}$ are invariant under the action of $\UU$. For instance, the states associated to coupled basis $\ket{j,j,L,M}$, or just $\ket{L,M}$ for short, have a Schmidt decomposition in terms of the decoupled basis $\ket{j,m_1} \ket{j,m_2}$. We write the explicit expressions for the states of $j=1$ and $3/2$ in Appendix~\ref{App.Cou.Dec}. In general, for a state $\ket{\psi} \in \IMA (\PP_{2j})$ to be connected by a $\UU$ transformation to a state $\ket{\phi} \in \IMA (\PP_{2j}^{\perp})$, their Schmidt numbers have to be equal. We apply this idea to the search for the optimal unitary operators, for $j=1$ and $3/2$. In particular, this approach led us to the identification of  $U_0$ for $j=3/2$, \ie, Eq.~\eqref{Eq.s3h2.U0}.
\subsection{\texorpdfstring{$N=2$ ($j=1$)}{Lg}}
Eq.~\eqref{ep.spin1} implies that the unitary transformations $\UU = U\otimes U$ that maximize $\ep$ are those that transformed one state from $\IMA (\PP_2)$ to $\IMA (\PP_0)$. Here, $\PP_0$ contains only one state given by (see Appendix~\ref{App.Cou.Dec})
\begin{equation}
    \ket{0,0} = \frac{1}{\sqrt{3}}\big( \ket{1}\ket{-1} - \ket{0}\ket{0} + \ket{-1}\ket{1} \big) ,
\end{equation}
with Schmidt numbers $(1,1,1)/\sqrt{3}$.
By direct inspection, we find that the state $\ket{\Psi} \in \IMA (\PP_2) $ given by
\begin{equation}
   \begin{aligned}       
   \ket{\Psi} = & \frac{1}{\sqrt{3}}\big( \sqrt{2} \ket{2,1} - \ket{2,-2} 
   \big)
   \\
   = & \frac{1}{\sqrt{3}}\big( \ket{1}\ket{0} - \ket{-1}\ket{-1} + \ket{0}\ket{1}  \big) 
      \end{aligned}
\end{equation} 
can be transformed, via $U_0'$ given in Eq.~\eqref{Eq.j1.U0p}, to $U_0' \otimes U_0' \ket{\Psi}= \ket{0,0} \in \IMA( \PP_0)$. 
Thus, $U_0'$ maximizes $\ep$.
\subsection{\texorpdfstring{$N=3$ ($j=3/2$)}{Lg}}
Similarly, Eq.~\eqref{ep.spin3h2} suggests that $\UU$ must transformed three vectors from $\IMA (\PP_3)$ to $\IMA (\PP_1)$ in order to optimize $\ep$. The $\ket{1,M}$ states have Schmidt numbers (see Appendix~\ref{App.Cou.Dec}) 
\begin{equation}
\begin{aligned}    
    \frac{1}{\sqrt{10}} \left( \sqrt{3} , \sqrt{3}, 2 , 0 \right) \quad & \text{ for } M=\pm 1 ,
\\
    \frac{1}{2\sqrt{5}} \left( 3 , 3, 1 , 1 \right) \quad & \text{ for } M=0 .
\end{aligned}
\end{equation}
We find that certain states in $\IMA (\PP_3)$ have the same Schmidt numbers. Moreover, the $\UU$ defined by Eq.~\eqref{Eq.s3h2.U0} effects the desired transformation
\begin{equation}
\begin{aligned}
    \UU \left(\sqrt{\frac{3}{5}}\ket{3,\mp 2}-\sqrt{\frac{2}{5}}\ket{3, \pm 3} \right) = &\ket{1, \pm 1}  ,
    \\
    \UU \ket{3,0} = & \ket{1,0} .
\end{aligned}
\end{equation}
\section{Conclusions and perspectives}
\label{Sec.Conclusions}
In this work, we have studied the entangling power of unitary operators acting on symmetric states of $N=2j$ qubits, which can be viewed as spin-$j$ states. 
The main differences with respect to the general non-symmetric case is that the set of product states are reduced to the SC states (see Eq.~\eqref{Eq.entangling.power}), and the set of unitary gates is reduced from SU$(2^N)$ to SU$(N+1)$ (See Sec.~\ref{Sec.Average} for more details). $\ep$ is reformulated as the inner product of two $(N+1)$-vectors~\eqref{Eq.EP.vec}. One vector, $\MM_q$, depends on the linear entanglement entropy $E_q$, while the other vector, $\NN$, is transformed by the unitary matrix $U$. The components of both vectors are SU$(2)$-invariant quantities, meaning that unitary transformations $U$, $U'$, differing only by left or right rotations, $U'=R_1 U R_2$, preserve them. Following this approach, several results and derivations are obtained, also raising new questions, which we summarize below.

We study in detail $\ep$ for states with a small number of qubits. Specifically, we identified the unitary gates that maximize $\ep(E_1,U)$ for $N=2$ and $N=3$, given by Eqs.~\eqref{Eq.j1.U0p} and \eqref{Eq.s3h2.U0}, respectively. Through numerical calculations, we also discovered two unitary gates, Eqs.~\eqref{Eq.U0q1j2}-\eqref{Eq.U0q2j2}, that are conjectured to maximize $\ep(E_q,U)$ for $N=4$ with $q=1,2$, respectively. These extremal unitaries possess highly symmetric entanglement distributions on the sphere (see Fig.~\ref{fig.Ent.Dis.Tod}) --- note that a similar characteristic is observed for the extremal spin states that maximize entanglement measures~\cite{PhysRevA.81.062347,Gir.Bra.Bra:10,Chr.etal:21,Gol.Kli.Gra.Leu.San:20}. Additionally, the hermitian matrices $\ED_q(U)$~\eqref{Eq.ED} of these extremal unitaries, associated with the entanglement distributions, display peculiar characteristics such as high point group symmetries and multiple degeneracies in their eigenspectra. Also notably, these extremal unitaries can be represented as linear combinations of at most two permutations matrices with complex entries. The $\ep$ reduces to an expression that contains the sum of the eigenvalues of $\ED_q$. We also point out  that $\ep(E_q ,U) =\ep(E_q ,U^{\dagger})$ holds true for $N=2$ and 3, but fails, in general, for a higher number of qubits. This symmetry of $\ep$ may be recast as time-reversal invariance, when $U$ is generated by a hamiltonian, $U=e^{-itH}$. 

We compute the average of $\ep(E_q,U)$ over unitary gates SU$(N+1)$ with respect to the Haar measure,  Eq.~\eqref{Eq.Average.Ep}, where we observe a difference with respect to the non-symmetric case Eq.~\eqref{Eq.Ave.Ep.NS}. By sampling Haar-uniform random unitaries, we find that most of their associated invariant vectors cluster near the mean value (see Table~\ref{Eq.Tab.Av.Ep}). These observations indicate that random unitary gates exhibit a statistical distribution with a narrow spread around the mean value. Further work could be done to derive the explicit statistical distribution of $\ep(E_q,U)$, as well as other variables such as the SU$(2)$-invariant components $p_L$~\eqref{Eq.bounds.components.P2j}. 

The vectors introduced in  our geometrical approach to the calculation of  $\ep$ have components associated to the subspace projectors $\PP_L$. Thus, finding the maximum of $\ep$ involves searching for the unitary gate that transforms a subspace into another. Additionally, unitary transformations do not alter the Schmidt decomposition of the states in $\PP_L$, suggesting that an alternative approach to maximizing $\ep$ is to examine the possible Schmidt decomposition  for the states spanning $\IMA(\PP_L)$. This approach proved useful in identifying the unitary gate that maximizes $\ep(E_1,U)$ for $N=3$.

Lastly, we remark that, for spin-$j$ pure states, the linear entanglement entropy $E_q$~\eqref{Eq.Ent.generalAB} coincides with the measure of anticoherence of order-$q$ based on the purity (see Eq.~(24) of Ref.~\cite{Bag.Mar:17}). Hence, the unitary gates with high (symmetric) entangling power correspond to those with high capacities to generate anticoherence in the SC states. Anticoherence for pure states is known to be a measure of non-classicality~\cite{Zim:06,Crann_2010,PhysRevLett.114.080401,Bag.Dam.Gir.Mar:15}. It is also known that anticoherent states are useful in quantum-enhanced metrology of rotations~\cite{Chr.Her:17,PhysRevA.98.032113,Chr.etal:21,Ser.Chr.Mar:24}. 

In summary, we have examined in detail the concept of entangling power for unitary matrices acting on symmetric multiqubit states, introducing reformulations in terms of inner products between vectors associated to SU$(2)$ invariants and transformation of subspaces of bipartite states among themselves. Additionally, the entanglement distribution of a unitary gate is associated with a linear combination of Husimi functions. These new perspectives could establish connections among other quantities relevant for quantum information theory that initially appear unrelated, including those relevant to the non-symmetric case. 
\section*{Acknowledgements}
ESE acknowledges support from the postdoctoral fellowship of the IPD-STEMA program of the University of Liège (Belgium). DMG acknowledges support from the F.R.S.-FNRS under the Excellence of Science (EOS) programme. CC acknowledges support from the UNAM-PAPIIT project IN112224.
JAM and DMG acknowledge support from the UNAM-PAPIIT project IN111122 (México).
\begin{appendix}
\section{Main equation}
\label{Appendix.Main.Equation}
Here, we derive Eq.~\eqref{Eq.Ent.q.first}. But first, we introduce the basis of $\HSs(\Hs^{(j)})$ defined by the \emph{multipole operators} $\left\{ T_{\si \mu}^{(j)} \right\}$, with $\si=0, \dots, 2j$ and $\mu=-\si, \dots ,\si$~\cite{Fan:53,Var.Mos.Khe:88} and where we omit the superindex when there is no possible confusion. The $T_{\si \mu}$ operators can be written in terms of the Clebsch-Gordan coefficients $C_{j_1 m_1 j_2 m_2}^{j m}$~\cite{Var.Mos.Khe:88} and they are orthonormal with respect to the HS scalar product
\begin{equation}
 \Tr \left( T_{\si_1 \mu_1}^{\dagger} T_{\si_2 \mu_2} \right) = \delta_{\si_1 \si_2} \delta_{\mu_1 \mu_2}   .
\end{equation}
A density matrix  $\rho \in \HSs (\Hs^{(j)})$, representing a quantum state, can always be expanded in the $\left\{ T_{\si \mu} \right\}$ basis
\begin{equation}
\label{rhoTLMexpansion}
    \rho = \sum_{\si=0}^{2j} \sum_{\mu=-\si}^{\si} \rho_{\si \mu} T_{\si \mu} =  \sum_{\si=0}^{2j} \bm{\rho}_{\si} \cdot \mathbf{T}_{\si}   , 
\end{equation}
where $\bm{\rho}_{\si} = \left( \rho_{\si \si } , \dots , \rho_{\si -\si} \right)$ with $\rho_{\si \mu} = \Tr (\rho T^{\dagger}_{\si \mu })$, and $\mathbf{T}_{\si}= \left( T_{\si \si} , \dots , T_{\si -\si} \right)$ is a vector of matrices. In particular, $T_{00}=(2j+1)^{-1/2} \mathds{1} $, and then $\rho_{00} = (2j+1)^{-1/2}$ for any mixed state. In the picture of the spin-$j$ states seeing as $2j$ spin-1/2 constituents, one can calculate the reduced density matrix $\rho_{q} = \Tr_{2j-q} \rho$, after tracing out $2j-q$ spins-1/2. Notably, the multipole expansion~\eqref{rhoTLMexpansion} of  $\rho_{q} $ has a simple expression in terms of the original state $\rho$~\cite{Ser.Bra:20,Den.Mar:22}
\begin{equation}
\label{Eq.Reduced.state}
   \left( \rho_q \right)_{\si \mu}= \frac{q!}{(2j)!} \sqrt{\frac{(2j-\si)!(2j+\si+1)!}{(q-\si)!(q+\si+1)!}} \rho_{\si \mu}  ,
\end{equation}
with $\si = 0 , \dots , q$ and $\mu=-\si , \dots , \si$.

Now, we calculate the entangling power~\eqref{Eq.Ent.generalAB} for a general spin-$j$ system. First, we write the general expression of the SC states, transformed by $U$,
\begin{equation}
    U \ket{j,\mathbf{n}}\bra{j,\mathbf{n}} U^{\dagger} = \frac{1}{\sqrt{2j+1}} T_{00}^{(j)} + \sum_{\si = 1 }^{2j} \bm{\rho}_{\si} \cdot \mathbf{T}_{\si}^{(j)}  , 
\end{equation}
with
\begin{equation}
    \rho_{\si \mu} = \bra{j,\mathbf{n}} U^{\dagger} T_{\si \mu}^{(j) \dagger} U \ket{j,\mathbf{n}}  .
\end{equation}
By tracing out the subsystem $B$, \ie, $2j-q$ spin-1/2 constituents, we obtain
\begin{equation}
\begin{aligned}    
    \rho_A = & \Tr_{2j-q} \left(
    U \ket{j,\mathbf{n}} \bra{j,\mathbf{n}} U^{\dagger} \right) \\
    = & \frac{1}{\sqrt{q+1}} T_{00}^{(q/2)} 
    \\
    + & \sum_{\si=1}^{q}
    \frac{q!}{(2j)!} \sqrt{\frac{(2j-\si)!(2j+\si+1)!}{(q-\si)!(q+\si+1)!}} \bm{\rho}_{\si} \cdot
    \mathbf{T}_{\si}^{(q/2)}
     ,
    \end{aligned}
\end{equation}
where we use Eq.~\eqref{Eq.Reduced.state}. We then get that 
\begin{equation}
\begin{aligned}    
    E_q(U\ket{j,\mathbf{n}}) = & 1 -  \frac{(q+1) (q!)^2}{q (2j)!^2} \times
\\
& 
 \sum_{\si=1}^q 
  \frac{(2j-\si)!(2j+\si+1)!}{ (q-\si)!(q+\si+1)!} \left|\bm{\rho}_{\si} \right|^2 ,
\\
= & 1 - \bra{j,\mathbf{n}} \otimes \bra{j,\mathbf{n}}
\UU^{\dagger} \MM_q \UU
\ket{j,\mathbf{n}} \otimes \ket{j,\mathbf{n}}
\end{aligned}
\end{equation}
where $\UU= U \otimes U$ as defined in the main text,
\begin{equation}
\TT_{\si} = \sum_{\mu = -\si}^{\si} T_{\si \mu} \otimes T_{\si \mu}^{\dagger} 
\end{equation}
and
\begin{equation}
\begin{aligned}
\label{Eq.Def.N.Mq}
%\TT_{e}
\MM_q = & \frac{(q+1)(q!)^2}{q(2j)!^2}
\sum_{\si=1}^q 
  \frac{(2j-\si)!(2j+\si+1)!}{ (q-\si)!(q+\si+1)!}
\TT_{\si}
\\
= &
\frac{q+1}{q}
\sum_{\si=1}^q 
\frac{\left\{
 \begin{array}{ccc}
     q/2 & q/2 & q  \\
     q/2 & q/2 & \si
 \end{array}
\right\} 
}{\left\{
 \begin{array}{ccc}
     j & j & 2j  \\
     j & j & \si
 \end{array}
\right\} 
}
  \TT_{\si}
.
\end{aligned}
\end{equation}
The $\TT_{\si}$'s can be written as a linear combination of the $\PP_L$ operators (see \textbf{Identity}~\ref{Identity.1} of Appendix~\ref{App.useful.id})
\begin{equation}
    \TT_{\si} = 
    (2\si+1) \sum_{L= 0}^{2j}  (-1)^{2j+L} 
\left\{ 
\begin{array}{ccc}
    j & j & L  \\
    j & j & \si   
\end{array}
\right\} \PP_{L}  .
\end{equation}
This result gives the expression~\eqref{Eq.Exp.Mq.PL} of $\MM_q$. Lastly, we use that $\ket{j,\mathbf{n}} \otimes \ket{j,\mathbf{n}} = \ket{2j,\mathbf{n}}$ to obtain Eq.~\eqref{Eq.Ent.q.first}.
\begin{table*}[t!]
    \centering
    \begin{tabular}{c | c | c }
      $j$
      & $\overrightarrow{\NN}$ in $\TT_{\si}$ basis  
      & $\overrightarrow{\MM_q} $ in $\TT_{\si}$ basis
      \\[7pt]
      \hline      
       1  
       & $\left(\frac{1}{3}, \frac{1}{2\sqrt{3}}  , \frac{1}{6\sqrt{5} }  \right)$ 
       & $  \left(0, 2\sqrt{3} ,0 \right)  $
       \\[7pt]
       3/2  
       & $\left(\frac{1}{4},\frac{3 \sqrt{3}}{20},\frac{1}{4 \sqrt{5}},\frac{1}{20 \sqrt{7}} \right)$ 
       & $ \left( 0,\frac{20}{3\sqrt{3}} ,0,0 \right)$
       \\[7pt]
       2  
       & $\left( \frac{1}{5},0,
       \frac{2}{5 \sqrt{3}},\frac{2}{7 \sqrt{5}},\frac{1}{10 \sqrt{7}},\frac{1}{210}
       \right)$ 
       & $\left\{ 
       \begin{array}{l}
        \left( 0, \frac{5\sqrt{3}}{2} ,0,0,0 \right)     \\[7pt]
         \left( 0, \frac{15 \sqrt{3}}{8},\frac{7 \sqrt{5}}{8},0,0 \right)
       \end{array}
        \right.$ 
       \\[15pt]
       5/2  
       & $\left( \frac{1}{6}, 0, \frac{5}{14 \sqrt{3}},\frac{5 \sqrt{5}}{84},\frac{5}{36 \sqrt{7}},\frac{1}{84},\frac{1}{252 \sqrt{11}} \right)$ 
       & $\left\{ 
       \begin{array}{l}
        \left( 0,\frac{14 \sqrt{3}}{5},0,0,0,0 \right)     \\[7pt]
         \left( 0,\frac{21 \sqrt{3}}{10},\frac{21}{5 \sqrt{5}},0,0,0 \right)
       \end{array}
        \right.$ 
    \end{tabular}
    \caption{Vectors of SU(2) invariants of the operators $\NN$ and $\MM_q$~\eqref{Eq.Def.N.Mq} in the $\TT_{\si}$  basis for the spin values $j=1 ,3/2 , 2,5/2$ and $q=1, \dots , \lfloor j \rfloor$. The vector $\overrightarrow{\MM}_q$ has $q$ non-zero entries.}
    \label{Table.ep.Tbasis}
\end{table*}
\section{The T operator basis}
\label{App.T.Basis}
\textbf{Identity~\ref{Identity.1}} of Appendix~\ref{App.useful.id} lets us write $\ep$~\eqref{Eq.Ent.q.first} as an inner product of vectors of SU(2) invariants associated to the $\TT_{\si}$ operators. The operator $\MM_q$ is written in Eq.~\eqref{Eq.Def.N.Mq}, while $\NN$ is expanded as
\begin{equation}
    \NN = \sum_{L= 0}^{2j} 
\left\{
 \begin{array}{ccc}
     j & j & 2j  \\
     j & j & L
 \end{array}
\right\} 
 \TT_{L} .
\end{equation}
The $\TT_{\sigma}$ operators are hermitian, orthogonal $
    \Tr (\TT_{\si} \TT_{\si'}) =(2\si+1) \delta_{\si \si'} 
$, and invariant under diagonal SU$(2)$ transformations in the double-copy space $\Hs^{(j) \otimes 2}$. Thus, and similarly as we did in Sec.~\ref{Sec.Reformulation}, we can associate to every operator $V \in \HSs(\Hs^{(j)\otimes 2})$ a $(2j+1)$-vector of $\mathrm{SU}(2)$ invariants
\begin{equation}
\overrightarrow{V}^{\TT}  = \Big( \Tr \left( V \tilde{\TT}_0 \right) , \dots , \Tr \left( V \tilde{\TT}_{\si} \right) , \dots , \Tr \left( V \tilde{\TT}_{2j} \right) \Big) ,
\end{equation}
with $\tilde{\TT}_{\si} = \TT_{\si} / \sqrt{2\si+1} $, and where the superscript denotes that the vector is written in the $\TT$ basis. We can follow the same procedure as in Sec.~\ref{Sec.Reformulation} to obtain the same reformulation of the $\ep$ as an inner product of two vectors given in Eqs.~\eqref{Eq.Trace.InnerP}-\eqref{Eq.EP.vec} but now expressed in the $\TT$ basis. We also have that each vector, now written in the $\TT$ basis, lies in a hyperplane, even after the action of $U$,
\begin{equation}
\label{Eq.hyperplane.R}
\begin{aligned} 
 \sum_{\si= 0}^{2j} \left(\MUU \overrightarrow{\NN} \right)_{\si}^{\TT} & = \sum_{\si= 0}^{2j} \Tr \left( \UU \NN \UU^{\dagger} \TT_{\sigma} \right)
 = 1 .
\\
\sum_{\si= 0}^{2j} \left(\MUU \overrightarrow{\MM}_q \right)_{\si}^{\TT}
& =   
   \frac{ 2(j+1)(2j+1)}{ (2 j-q+1)} .
\end{aligned}
\end{equation}
We prove these results in Identity~\ref{Identity.3} of Appendix~\ref{App.useful.id}.
Since $\TT_0$ is propotional to the identity, $\left( \MUU \overrightarrow{V} \right)^{\TT}_0 = \left( \overrightarrow{V} \right)^{\TT}_0$. In particular $\left( \overrightarrow{\NN} \right)^{\TT}_0 = 1/(2j+1)$ and $\left( \overrightarrow{\MM}_q \right)^{\TT}_0 = 0$.  We write $\NN$ and $\MM_q$ for $j=1,3/2,2$ and $5/2$ in the $\TT$ basis in Table~\ref{Table.ep.Tbasis}. The properties of the 6j-symbols show that the components of $\overrightarrow{\NN}$ and $\overrightarrow{\MM}_q$ are always positive in the $\TT$ basis~\cite{Var.Mos.Khe:88}. However, the disadvantage of the $\TT$ basis is that a $\UU$ transformation combines all the $\TT$ components, not only half of them as in the $\PP$ basis.
\begin{widetext}
\section{Useful identities}
\label{App.useful.id}
Here, we prove several identities used throughout the paper. 
\begin{identity} 
\label{Identity.1}
There is a linear transformation between the operators $\TT_{\si}$ and $\PP_L$
\begin{equation}
    \TT_{\si} = 
    (2\si+1) \sum_{L= 0}^{2j}  (-1)^{2j+L} 
\left\{ 
\begin{array}{ccc}
    j & j & L  \\
    j & j & \si   
\end{array}
\right\} \PP_{L} 
 ,
\quad \quad 
   \mathcal{P}_{L} = (-1)^{2j + L} (2L+1) \sum_{\si=0}^{2j}     \left\{ 
\begin{array}{ccc}
    j & j & \si \\
    j & j & L   
\end{array}
\right\}
\TT_{\si}  
 .
\end{equation}    
\end{identity}
\emph{Proof.}
\begin{equation}
    \begin{aligned}
    \TT_{\si} = &
         \sum_{\mu =-\si}^{\si} T_{\si \mu} \otimes T_{\si \mu}^{\dagger}
\\
= & 
    \left( \frac{2\si+1}{2j+1} \right)\sum_{\mu} \sum_{\substack{m,m'\\ n,n'}} C^{jm}_{jm',\si \mu } C^{jn}_{jn',\si \mu } \ket{j,m}\bra{j,m'} \otimes \ket{j,n'}\bra{j,n}
    \\
= &      
\left( \frac{2\si+1}{2j+1} \right)\sum_{\mu} \sum_{\substack{m,m'\\ n,n'}} \sum_{\substack{L_1 , M_1\\ L_2 , M_2}} C^{jm}_{jm',\si \mu} C^{jn}_{jn',\si \mu} C^{L_1 M_1}_{jm,jn'} C^{L_2 M_2}_{jm',jn} \ket{j,j,L_1,M_1} \bra{j,j,L_2,M_2}
\\
= &      
\left( \frac{2\si+1}{2j+1}  \right) (-1)^{\si} \sum_{\substack{L_1 , M_1\\ L_2 , M_2}} \sum_{M} \sum_{\substack{m,m'\\ n,n'}} 
C^{L_1 M_1}_{jm,jn'} C^{jm}_{jm',\si \mu} C^{L_2 M_2}_{jm',jn} C^{jn}_{\si \mu,jn'}   \ket{j,j,L_1,M_1} \bra{j,j, L_2,M_2}
\\
= & 
\left( 2\si+1 \right) \sum_{\substack{L_1 , M_1}} (-1)^{2j+L_1}
\left\{ 
\begin{array}{ccc}
    j & j & L_1  \\
    j & j & \si   
\end{array}
\right\} \ket{L_1,M_1}_C \bra{L_1,M_1}_C 
\\
= & (2\si+1) \sum_{L= 0}^{2j}  (-1)^{2j+L} 
\left\{ 
\begin{array}{ccc}
    j & j & L  \\
    j & j & \si   
\end{array}
\right\} \PP_{L}  ,
    \end{aligned}
\end{equation}
where we use the decoupled and coupled bases of the angular momentum~\eqref{Eq.coupled.basis} and the definition of the 6j symbol (see p.~291, Eq.~(8) of Ref.~\cite{Var.Mos.Khe:88}). Lastly, we invert the previous equation with the orthogonality of the 6j symbols (See page 291 or Ref.~\cite{Var.Mos.Khe:88} for the general formula)
\begin{equation}
\label{Eq.Ort.6j}
    (2\si+1) \sum_{L=0}^{2j} (2L+1)
    \left\{ 
\begin{array}{ccc}
    j & j & L  \\
    j & j & \si   
\end{array}
\right\}
    \left\{ 
\begin{array}{ccc}
    j & j & L  \\
    j & j & \si'   
\end{array}
\right\} = \delta_{\si \si'}  ,
\end{equation}
to write $\PP_L$ in terms of the operators $\TT_{\si}$. $\square$

\begin{identity}
\label{Identity.2} 
For any unitary transformation $\UU= U\otimes U$ and operator $V= \sum_{K=0}^{2j} v_K \PP_K$, it holds that 
\begin{equation}
\label{Eq.Identity.2}
    \sum_{\substack{L=0 \\ 2j \equiv L +\delta \Mod{2}}}^{2j} \hspace{-0.5cm} \Tr \left( \UU V \UU^{\dagger} \PP_{L} \right) 
    = \sum_{\substack{L=0 \\ 2j \equiv L +\delta \Mod{2}}}^{2j} \hspace{-0.5cm} \Tr \left(  V \PP_{L} \right) ,
\end{equation}
for $\delta=0$ and 1, or when $L$ runs over all the indices. In particular,
\begin{equation}
    \sum_{\substack{L=0 \\ K \equiv L \Mod 2}}^{2j} \hspace{-0.5cm} \Tr \left( \UU \PP_{L} \UU^{\dagger} \PP_{K} \right) = 2K+1 
 , \quad \quad 
    \sum_{\substack{L=0 \\ 2j \equiv L \Mod{2}}}^{2j} \hspace{-0.5cm} \Tr \left( \UU \PP_{L} \UU^{\dagger} \MM_q \right)  = \frac{(j+1)(2j+1)}{2j+1-q} 
    .
\end{equation}    
\end{identity}
\emph{Proof.} By resolution of the unity of the $\PP_L$ operators, we have that
\begin{equation}
 \sum_{L=0}^{2j} \Tr \left( 
 \UU \PP_L \UU^{\dagger} \PP_K \right)  
 =
 \Tr \left( \PP_K \right)  
 = 2K+1 .
\end{equation}
Now, the unitary transformations $\UU = U\otimes U$ preserve the orthogonality between the projectors of different parity
\begin{equation}
    \Tr (\UU \PP_L \UU^{\dagger} \PP_K) = 0  , \quad \text{for } L \equiv K +1 \Mod{2}  ,
\end{equation}
because they preserve the permutation exchange of the states $\ket{\Psi }\in  \Hs^{(j) \otimes 2}$. Thus, any operator $V= \sum_{K=0}^{2j} v_K \PP_K$ can be split in the components with $K$ odd and even, and these components do not mixed after a transformation $\UU$. This proves~\eqref{Eq.Identity.2}. In particular,
\begin{equation}
 \sum_{L=0}^{2j} \Tr \left( 
 \UU \PP_L \UU^{\dagger} \PP_K \right)  
 =
 \sum_{\substack{L=0 \\ K \equiv L  \Mod{2}}}^{2j} \Tr \left( 
 \UU \PP_L \UU^{\dagger} \PP_K \right)  
 =
 \Tr \left( \PP_K \right)  
 = 2K+1 .
\end{equation}
For $\MM_q$, we have that
\begin{equation}
\label{Eq.Sum}
     \sum_{\substack{L=0 \\ 2j \equiv L +\delta \Mod{2}}}^{2j} \hspace{-0.3cm} 
     \Tr \left( \UU \MM_q \UU^{\dagger} \PP_L \right) 
=
\frac{1}{2} \sum_{L=0 }^{2j} 
\left[ 1+ (-1)^{2j+L+\delta} \right]
\Tr \left( \MM_q \PP_L \right) 
= 
(-1)^{\delta} \frac{(j+1)(2j+1)}{2j+1-q}
,
\end{equation}
where we use identities of the 6j symbols~\cite{Var.Mos.Khe:88}. $\square$
\begin{identity}
\label{Identity.3}
Any hermitian operator $ V = \sum_{\si= 0}^{2j} v_{\si} \TT_{\si} $ satisfies
\begin{equation}   \label{Norm1.invariance.R.general}
    \sum_{\si= 0}^{2j} \Tr \left( \UU V \UU^{\dagger} \TT_{\si} \right) = \sum_{\si= 0}^{2j} \Tr \left(  V \TT_{\si} \right) ,
\end{equation}
where $\UU = U\otimes U$ and $U$ a unitary operator. In particular, 
\begin{equation}
    \begin{aligned}        
        \label{Id.a}
        \sum_{\si=0}^{2j} \Tr \left( \MM_q \TT_{\si} \right) 
        = \frac{ 2(j+1)(2j+1)}{ (2 j-q+1)}
    , \quad \quad 
        \sum_{\si=0}^{2j} \Tr \left( \NN \TT_{\si} \right)
=  \frac{2j}{2j+1}
        .
    \end{aligned}
    \end{equation}
\end{identity}
\emph{Proof.} The following caclulation will prove convenient later on,
\begin{equation}
    \begin{aligned}
        \sum_{\si=0}^{2j} \TT_{\si} & = \sum_{\si =0 }^{2j} \sum_{\mu =-\si}^{\si} \sum_{\substack{m_1 , m_2 \\ n_1 , n_2}} (-1)^{2j-m_2 -n_2} C_{j m_1 j -m_2}^{\si \mu} C_{j n_1 j -n_2}^{\si \mu} \ket{j , m_1 , j , n_2} \bra{j, m_2 , j , n_1 } 
\\
& = \sum_{\substack{m_1 , m_2 \\ n_1 , n_2}} (-1)^{2j-m_2 -n_2} \delta_{m_1, n_1} \delta_{m_2 n_2} \ket{j , m_1 , j , n_2} \bra{j, m_2 , j , n_1 }
\\
& = 
\sum_{\substack{m_1 , m_2 }} \ket{j , m_1 , j , m_2} \bra{j, m_2 , j , m_1 }
    \end{aligned}
\end{equation}
where we use that $2(j-m)$ is always an even integer number. 
Now, let us prove the result for the $\TT_{\si}$ operators
\begin{equation}
        \sum_{\si= 0}^{2j} \Tr \left( \UU^{\dagger} \TT_{\tau} \UU \TT_{\si} \right) 
         = 
        \sum_{\nu = -\tau}^{\tau} \sum_{m_1, m_2} \bra{j,m_2} U^{\dagger} T_{\tau \nu} U \ket{j,m_1 } \bra{j,m_1 } U^{\dagger} T_{\tau \nu}^{\dagger} U \ket{j,m_2} 
        = 2\tau + 1 =
        \sum_{\si= 0}^{2j} \Tr \left(  \TT_{\tau} \TT_{\si} \right) 
         ,
\end{equation}
The first equality will be valid for any $V$ by linearity. In particular, we get for $V= \NN$ that
   \begin{equation}
    \begin{aligned}  
        \sum_{\si=0}^{2j} \Tr \left( \NN \TT_{\si} \right) = \sum_{\si=0}^{2j} \sum_{L= 1}^{2j} 
        \left\{ 
\begin{array}{ccc}
    j & j & 2j  \\
    j & j & L   
\end{array}
\right\}
\Tr \left( \TT_{L} \TT_{\si} \right) 
= 
        \sum_{L= 1}^{2j} 
        (2L + 1)
        \left\{ 
\begin{array}{ccc}
    j & j & 2j  \\
    j & j & L   
\end{array}
\right\} 
        = \frac{2j}{2j+1}
         ,
    \end{aligned}
    \end{equation}
where we use the following identities of the 6j symbol~\cite{Var.Mos.Khe:88}
\begin{equation}
\sum_{L=0}^{2j} (2L+1)         \left\{ 
\begin{array}{ccc}
    j & j & 2j  \\
    j & j & L
\end{array}
\right\}
= 1  , 
\quad
\left\{ 
\begin{array}{ccc}
    j & j & 2j  \\
    j & j & 0   
\end{array}
\right\}
= \frac{1}{2j+1}  .  \quad 
\end{equation}
Similarly, 
\begin{equation}
        \sum_{\si=0}^{2j} \Tr \left( \MM_q \TT_{\si} \right) =
        \frac{q+1}{q}
\sum_{\tau=1}^q (2\tau+1)
\frac{\left\{
 \begin{array}{ccc}
     q/2 & q/2 & q  \\
     q/2 & q/2 & \tau
 \end{array}
\right\} 
}{\left\{
 \begin{array}{ccc}
     j & j & 2j  \\
     j & j & \tau
 \end{array}
\right\} 
} 
=
\frac{ 2(j+1)(2j+1)}{ 2j+1-q } . \quad \quad \square
\end{equation}
\section{Proof of Eq.\texorpdfstring{~\eqref{Eq.Average.Ep}}{Lg}}
\label{App.Average.EP}
We start with the following identity regarding the integration over the SU$(d)$ unitary matrices and where $A$ is a $d^2 \times d^2$ matrix (see \textbf{Proposition 3.9} in~\cite{zhang2014matrix})
\begin{equation}
    \int \UU A \UU^{\dagger} \diff \mu (U) = \left[ \frac{\Tr (A)}{d^2-1} - \frac{\Tr (AF)}{d(d^2-1)} \right] \mathds{1}_{d^2} -
    \left[ \frac{\Tr (A)}{d(d^2-1)} - \frac{\Tr (AF)}{d^2-1} \right] F
\end{equation}
where $F \ket{\psi_1} \ket{\psi_2} = \ket{\psi_2} \ket{\psi_1} $ is the swap operator. In particular $F \PP_{L} = \PP_{L} F = (-1)^{2j-L} \PP_L$. Thus,
\begin{equation}
    \begin{aligned}
     \int \ep (E_q ,U) \diff \mu (U) 
     = & 1 - \Tr \left( \left[ \int \UU \NN \UU^{\dagger} \diff \mu (U) \right]  \MM_q \right)
     = 1 - \frac{1}{d(d+1)} \left[ \Tr \left(\MM_q \right) + \Tr \left( F \MM_q \right) \right] .
    \end{aligned}
\end{equation}
We add the resolution of the unity $\mathds{1}_{d^2} = \sum_{L=0}^{2j} \PP_L$ to use Eq.~\eqref{Eq.Sum} in the last equation
\begin{equation}
    \int \ep (E_q ,U) \diff \mu (U)  = 1- \frac{2(j+1)(2j+1)}{d(d+1)(2j+1-q)} = \frac{2j-q}{2j+1-q}.
\end{equation}
\section{Spin states in the coupled and decoupled bases}
\label{App.Cou.Dec}
Here, we write the coupled basis $\ket{L,M} \equiv \ket{j_1, j_2 ,L , M}$ in terms of the decoupled basis $\ket{m_1} \ket{m_2} \equiv \ket{j_1,m_1}\ket{j_2 m_2}$. For $j=1$, we have
\begin{equation}
\begin{aligned}
\ket{2,\pm 2} & = \ket{\pm 1}\ket{\pm 1}  ,
    \\
    \ket{2, \pm 1} & = \frac{1}{\sqrt{2}}\big( \ket{\pm 1}\ket{0} + \ket{0}\ket{ \pm 1} \big)  ,
    \\
    \ket{2,0} & = \frac{1}{\sqrt{6}}\big( \ket{1}\ket{-1} + 2\ket{0}\ket{0} + \ket{-1}\ket{1} \big)  ,
    \\
    \ket{0,0} & = \frac{1}{\sqrt{3}}\big( \ket{1}\ket{-1} - \ket{0}\ket{0} + \ket{-1}\ket{1} \big)  ,   
\end{aligned}
\end{equation}
And for $j=3/2$, it reads
\begin{equation}
\begin{aligned}
\ket{3, \pm 3 }= & \left| \pm \frac{3}{2} \right\rangle  \left| \pm \frac{3}{2} \right\rangle 
 ,
\\
\ket{3, \pm 2 }= & \frac{1}{\sqrt{2}} \left( \left| \pm \frac{3}{2} \right\rangle  \left| \pm \frac{1}{2} \right\rangle + \left| \pm \frac{1}{2} \right\rangle  \left| \pm \frac{3}{2} \right\rangle \right)
 ,
\\
\ket{3, \pm 1 }= & \frac{1}{\sqrt{5}} \left( \left| \pm \frac{3}{2} \right\rangle  \left| \mp \frac{1}{2} \right\rangle +
  \left| \mp \frac{1}{2} \right\rangle
  \left| \pm \frac{3}{2} \right\rangle
 +\sqrt{3} \left| \pm \frac{1}{2} \right\rangle  \left| \pm \frac{1}{2} \right\rangle \right)
 ,
\\
\ket{3, 0 }= & \frac{1}{2\sqrt{5}} \left( \left| \frac{3}{2} \right\rangle  \left| -\frac{3}{2} \right\rangle +
\left| -\frac{3}{2} \right\rangle  \left| \frac{3}{2} \right\rangle + 3 \left| \frac{1}{2} \right\rangle  \left| - \frac{1}{2} \right\rangle
 + 3 \left| -\frac{1}{2} \right\rangle  \left|  \frac{1}{2} \right\rangle \right)
 ,
\end{aligned}
\end{equation}
\begin{equation}
    \begin{aligned}
        \ket{1, \pm 1} = & \frac{1}{\sqrt{10}} \left(
\sqrt{3}  \left| \pm \frac{3}{2} \right\rangle  \left| \mp \frac{1}{2} \right\rangle
+
\sqrt{3} \left| \mp \frac{1}{2} \right\rangle
 \left| \pm \frac{3}{2} \right\rangle   -2 \left| \pm \frac{1}{2} \right\rangle
 \left| \pm \frac{1}{2} \right\rangle \right)
 ,
\\
\ket{1, 0 }= & \frac{1}{2\sqrt{5}} \left( 3 \left| \frac{3}{2} \right\rangle  \left| -\frac{3}{2} \right\rangle + 3
\left| -\frac{3}{2} \right\rangle  \left| \frac{3}{2} \right\rangle - \left| \frac{1}{2} \right\rangle  \left| - \frac{1}{2} \right\rangle
 - \left| -\frac{1}{2} \right\rangle  \left|  \frac{1}{2} \right\rangle \right)
 ,
    \end{aligned}
\end{equation}
\end{widetext}
\end{appendix}
\bibliographystyle{apsrev4-2}
\bibliography{Refs_entangling_power}

\begin{thebibliography}{57}%
\makeatletter
\providecommand \@ifxundefined [1]{%
 \@ifx{#1\undefined}
}%
\providecommand \@ifnum [1]{%
 \ifnum #1\expandafter \@firstoftwo
 \else \expandafter \@secondoftwo
 \fi
}%
\providecommand \@ifx [1]{%
 \ifx #1\expandafter \@firstoftwo
 \else \expandafter \@secondoftwo
 \fi
}%
\providecommand \natexlab [1]{#1}%
\providecommand \enquote  [1]{``#1''}%
\providecommand \bibnamefont  [1]{#1}%
\providecommand \bibfnamefont [1]{#1}%
\providecommand \citenamefont [1]{#1}%
\providecommand \href@noop [0]{\@secondoftwo}%
\providecommand \href [0]{\begingroup \@sanitize@url \@href}%
\providecommand \@href[1]{\@@startlink{#1}\@@href}%
\providecommand \@@href[1]{\endgroup#1\@@endlink}%
\providecommand \@sanitize@url [0]{\catcode `\\12\catcode `\$12\catcode
  `\&12\catcode `\#12\catcode `\^12\catcode `\_12\catcode `\%12\relax}%
\providecommand \@@startlink[1]{}%
\providecommand \@@endlink[0]{}%
\providecommand \url  [0]{\begingroup\@sanitize@url \@url }%
\providecommand \@url [1]{\endgroup\@href {#1}{\urlprefix }}%
\providecommand \urlprefix  [0]{URL }%
\providecommand \Eprint [0]{\href }%
\providecommand \doibase [0]{https://doi.org/}%
\providecommand \selectlanguage [0]{\@gobble}%
\providecommand \bibinfo  [0]{\@secondoftwo}%
\providecommand \bibfield  [0]{\@secondoftwo}%
\providecommand \translation [1]{[#1]}%
\providecommand \BibitemOpen [0]{}%
\providecommand \bibitemStop [0]{}%
\providecommand \bibitemNoStop [0]{.\EOS\space}%
\providecommand \EOS [0]{\spacefactor3000\relax}%
\providecommand \BibitemShut  [1]{\csname bibitem#1\endcsname}%
\let\auto@bib@innerbib\@empty
%</preamble>
\bibitem [{\citenamefont {Horodecki}\ \emph {et~al.}(2009)\citenamefont
  {Horodecki}, \citenamefont {Horodecki}, \citenamefont {Horodecki},\ and\
  \citenamefont {Horodecki}}]{RevModPhys.81.865}%
  \BibitemOpen
  \bibfield  {author} {\bibinfo {author} {\bibfnamefont {R.}~\bibnamefont
  {Horodecki}}, \bibinfo {author} {\bibfnamefont {P.}~\bibnamefont
  {Horodecki}}, \bibinfo {author} {\bibfnamefont {M.}~\bibnamefont
  {Horodecki}},\ and\ \bibinfo {author} {\bibfnamefont {K.}~\bibnamefont
  {Horodecki}},\ }\href {https://doi.org/10.1103/RevModPhys.81.865} {\bibfield
  {journal} {\bibinfo  {journal} {Rev. Mod. Phys.}\ }\textbf {\bibinfo {volume}
  {81}},\ \bibinfo {pages} {865} (\bibinfo {year} {2009})}\BibitemShut
  {NoStop}%
\bibitem [{\citenamefont {Nielsen}\ and\ \citenamefont
  {Chuang}(2011)}]{Nie.Chu:11}%
  \BibitemOpen
  \bibfield  {author} {\bibinfo {author} {\bibfnamefont {M.~A.}\ \bibnamefont
  {Nielsen}}\ and\ \bibinfo {author} {\bibfnamefont {I.~L.}\ \bibnamefont
  {Chuang}},\ }\href@noop {} {\emph {\bibinfo {title} {Quantum {Computation}
  and {Quantum} {Information}: 10th {Anniversary} {Edition}}}},\ \bibinfo
  {edition} {10th}\ ed.\ (\bibinfo  {publisher} {Cambridge University Press},\
  \bibinfo {address} {New York, NY, USA},\ \bibinfo {year} {2011})\BibitemShut
  {NoStop}%
\bibitem [{\citenamefont {Bengtsson}\ and\ \citenamefont
  {\.{Z}yczkowski}(2017)}]{Ben.Zyc:17}%
  \BibitemOpen
  \bibfield  {author} {\bibinfo {author} {\bibfnamefont {I.}~\bibnamefont
  {Bengtsson}}\ and\ \bibinfo {author} {\bibfnamefont {K.}~\bibnamefont
  {\.{Z}yczkowski}},\ }\href@noop {} {\emph {\bibinfo {title} {Geometry of
  Quantum States (2nd {E}d.)}}}\ (\bibinfo  {publisher} {Cambridge University
  Press},\ \bibinfo {year} {2017})\BibitemShut {NoStop}%
\bibitem [{\citenamefont {D\"ur}\ \emph {et~al.}(2001)\citenamefont {D\"ur},
  \citenamefont {Vidal}, \citenamefont {Cirac}, \citenamefont {Linden},\ and\
  \citenamefont {Popescu}}]{PhysRevLett.87.137901}%
  \BibitemOpen
  \bibfield  {author} {\bibinfo {author} {\bibfnamefont {W.}~\bibnamefont
  {D\"ur}}, \bibinfo {author} {\bibfnamefont {G.}~\bibnamefont {Vidal}},
  \bibinfo {author} {\bibfnamefont {J.~I.}\ \bibnamefont {Cirac}}, \bibinfo
  {author} {\bibfnamefont {N.}~\bibnamefont {Linden}},\ and\ \bibinfo {author}
  {\bibfnamefont {S.}~\bibnamefont {Popescu}},\ }\href
  {https://doi.org/10.1103/PhysRevLett.87.137901} {\bibfield  {journal}
  {\bibinfo  {journal} {Phys. Rev. Lett.}\ }\textbf {\bibinfo {volume} {87}},\
  \bibinfo {pages} {137901} (\bibinfo {year} {2001})}\BibitemShut {NoStop}%
\bibitem [{\citenamefont {Wolf}\ \emph {et~al.}(2003)\citenamefont {Wolf},
  \citenamefont {Eisert},\ and\ \citenamefont
  {Plenio}}]{PhysRevLett.90.047904}%
  \BibitemOpen
  \bibfield  {author} {\bibinfo {author} {\bibfnamefont {M.~M.}\ \bibnamefont
  {Wolf}}, \bibinfo {author} {\bibfnamefont {J.}~\bibnamefont {Eisert}},\ and\
  \bibinfo {author} {\bibfnamefont {M.~B.}\ \bibnamefont {Plenio}},\ }\href
  {https://doi.org/10.1103/PhysRevLett.90.047904} {\bibfield  {journal}
  {\bibinfo  {journal} {Phys. Rev. Lett.}\ }\textbf {\bibinfo {volume} {90}},\
  \bibinfo {pages} {047904} (\bibinfo {year} {2003})}\BibitemShut {NoStop}%
\bibitem [{\citenamefont {Xiong}\ \emph {et~al.}(2011)\citenamefont {Xiong},
  \citenamefont {Lu},\ and\ \citenamefont {Wang}}]{Xiong_2012}%
  \BibitemOpen
  \bibfield  {author} {\bibinfo {author} {\bibfnamefont {H.-N.}\ \bibnamefont
  {Xiong}}, \bibinfo {author} {\bibfnamefont {X.-M.}\ \bibnamefont {Lu}},\ and\
  \bibinfo {author} {\bibfnamefont {X.}~\bibnamefont {Wang}},\ }\href
  {https://doi.org/10.1088/0953-4075/45/1/015501} {\bibfield  {journal}
  {\bibinfo  {journal} {Journal of Physics B: Atomic, Molecular and Optical
  Physics}\ }\textbf {\bibinfo {volume} {45}},\ \bibinfo {pages} {015501}
  (\bibinfo {year} {2011})}\BibitemShut {NoStop}%
\bibitem [{\citenamefont {Kraus}\ and\ \citenamefont
  {Cirac}(2001)}]{PhysRevA.63.062309}%
  \BibitemOpen
  \bibfield  {author} {\bibinfo {author} {\bibfnamefont {B.}~\bibnamefont
  {Kraus}}\ and\ \bibinfo {author} {\bibfnamefont {J.~I.}\ \bibnamefont
  {Cirac}},\ }\href {https://doi.org/10.1103/PhysRevA.63.062309} {\bibfield
  {journal} {\bibinfo  {journal} {Phys. Rev. A}\ }\textbf {\bibinfo {volume}
  {63}},\ \bibinfo {pages} {062309} (\bibinfo {year} {2001})}\BibitemShut
  {NoStop}%
\bibitem [{\citenamefont {Zhang}\ \emph {et~al.}(2003)\citenamefont {Zhang},
  \citenamefont {Vala}, \citenamefont {Sastry},\ and\ \citenamefont
  {Whaley}}]{PhysRevA.67.042313}%
  \BibitemOpen
  \bibfield  {author} {\bibinfo {author} {\bibfnamefont {J.}~\bibnamefont
  {Zhang}}, \bibinfo {author} {\bibfnamefont {J.}~\bibnamefont {Vala}},
  \bibinfo {author} {\bibfnamefont {S.}~\bibnamefont {Sastry}},\ and\ \bibinfo
  {author} {\bibfnamefont {K.~B.}\ \bibnamefont {Whaley}},\ }\href
  {https://doi.org/10.1103/PhysRevA.67.042313} {\bibfield  {journal} {\bibinfo
  {journal} {Phys. Rev. A}\ }\textbf {\bibinfo {volume} {67}},\ \bibinfo
  {pages} {042313} (\bibinfo {year} {2003})}\BibitemShut {NoStop}%
\bibitem [{\citenamefont {Kalsi}\ \emph {et~al.}(2022)\citenamefont {Kalsi},
  \citenamefont {Romito},\ and\ \citenamefont {Schomerus}}]{Kalsi_2022}%
  \BibitemOpen
  \bibfield  {author} {\bibinfo {author} {\bibfnamefont {T.}~\bibnamefont
  {Kalsi}}, \bibinfo {author} {\bibfnamefont {A.}~\bibnamefont {Romito}},\ and\
  \bibinfo {author} {\bibfnamefont {H.}~\bibnamefont {Schomerus}},\ }\href
  {https://doi.org/10.1088/1751-8121/ac71e8} {\bibfield  {journal} {\bibinfo
  {journal} {Journal of Physics A: Mathematical and Theoretical}\ }\textbf
  {\bibinfo {volume} {55}},\ \bibinfo {pages} {264009} (\bibinfo {year}
  {2022})}\BibitemShut {NoStop}%
\bibitem [{\citenamefont {Tang}\ \emph {et~al.}(2023)\citenamefont {Tang},
  \citenamefont {Connelly}, \citenamefont {Warren}, \citenamefont {Zhuang},
  \citenamefont {Economou},\ and\ \citenamefont
  {Barnes}}]{PhysRevApplied.19.044094}%
  \BibitemOpen
  \bibfield  {author} {\bibinfo {author} {\bibfnamefont {H.~L.}\ \bibnamefont
  {Tang}}, \bibinfo {author} {\bibfnamefont {K.}~\bibnamefont {Connelly}},
  \bibinfo {author} {\bibfnamefont {A.}~\bibnamefont {Warren}}, \bibinfo
  {author} {\bibfnamefont {F.}~\bibnamefont {Zhuang}}, \bibinfo {author}
  {\bibfnamefont {S.~E.}\ \bibnamefont {Economou}},\ and\ \bibinfo {author}
  {\bibfnamefont {E.}~\bibnamefont {Barnes}},\ }\href
  {https://doi.org/10.1103/PhysRevApplied.19.044094} {\bibfield  {journal}
  {\bibinfo  {journal} {Phys. Rev. Appl.}\ }\textbf {\bibinfo {volume} {19}},\
  \bibinfo {pages} {044094} (\bibinfo {year} {2023})}\BibitemShut {NoStop}%
\bibitem [{\citenamefont {Caruso}\ \emph {et~al.}(2010)\citenamefont {Caruso},
  \citenamefont {Chin}, \citenamefont {Datta}, \citenamefont {Huelga},\ and\
  \citenamefont {Plenio}}]{PhysRevA.81.062346}%
  \BibitemOpen
  \bibfield  {author} {\bibinfo {author} {\bibfnamefont {F.}~\bibnamefont
  {Caruso}}, \bibinfo {author} {\bibfnamefont {A.~W.}\ \bibnamefont {Chin}},
  \bibinfo {author} {\bibfnamefont {A.}~\bibnamefont {Datta}}, \bibinfo
  {author} {\bibfnamefont {S.~F.}\ \bibnamefont {Huelga}},\ and\ \bibinfo
  {author} {\bibfnamefont {M.~B.}\ \bibnamefont {Plenio}},\ }\href
  {https://doi.org/10.1103/PhysRevA.81.062346} {\bibfield  {journal} {\bibinfo
  {journal} {Phys. Rev. A}\ }\textbf {\bibinfo {volume} {81}},\ \bibinfo
  {pages} {062346} (\bibinfo {year} {2010})}\BibitemShut {NoStop}%
\bibitem [{\citenamefont {Zanardi}\ \emph {et~al.}(2000)\citenamefont
  {Zanardi}, \citenamefont {Zalka},\ and\ \citenamefont
  {Faoro}}]{PhysRevA.62.030301}%
  \BibitemOpen
  \bibfield  {author} {\bibinfo {author} {\bibfnamefont {P.}~\bibnamefont
  {Zanardi}}, \bibinfo {author} {\bibfnamefont {C.}~\bibnamefont {Zalka}},\
  and\ \bibinfo {author} {\bibfnamefont {L.}~\bibnamefont {Faoro}},\ }\href
  {https://doi.org/10.1103/PhysRevA.62.030301} {\bibfield  {journal} {\bibinfo
  {journal} {Phys. Rev. A}\ }\textbf {\bibinfo {volume} {62}},\ \bibinfo
  {pages} {030301} (\bibinfo {year} {2000})}\BibitemShut {NoStop}%
\bibitem [{\citenamefont {Nielsen}\ \emph {et~al.}(2003)\citenamefont
  {Nielsen}, \citenamefont {Dawson}, \citenamefont {Dodd}, \citenamefont
  {Gilchrist}, \citenamefont {Mortimer}, \citenamefont {Osborne}, \citenamefont
  {Bremner}, \citenamefont {Harrow},\ and\ \citenamefont
  {Hines}}]{PhysRevA.67.052301}%
  \BibitemOpen
  \bibfield  {author} {\bibinfo {author} {\bibfnamefont {M.~A.}\ \bibnamefont
  {Nielsen}}, \bibinfo {author} {\bibfnamefont {C.~M.}\ \bibnamefont {Dawson}},
  \bibinfo {author} {\bibfnamefont {J.~L.}\ \bibnamefont {Dodd}}, \bibinfo
  {author} {\bibfnamefont {A.}~\bibnamefont {Gilchrist}}, \bibinfo {author}
  {\bibfnamefont {D.}~\bibnamefont {Mortimer}}, \bibinfo {author}
  {\bibfnamefont {T.~J.}\ \bibnamefont {Osborne}}, \bibinfo {author}
  {\bibfnamefont {M.~J.}\ \bibnamefont {Bremner}}, \bibinfo {author}
  {\bibfnamefont {A.~W.}\ \bibnamefont {Harrow}},\ and\ \bibinfo {author}
  {\bibfnamefont {A.}~\bibnamefont {Hines}},\ }\href
  {https://doi.org/10.1103/PhysRevA.67.052301} {\bibfield  {journal} {\bibinfo
  {journal} {Phys. Rev. A}\ }\textbf {\bibinfo {volume} {67}},\ \bibinfo
  {pages} {052301} (\bibinfo {year} {2003})}\BibitemShut {NoStop}%
\bibitem [{\citenamefont {Jonnadula}\ \emph {et~al.}(2017)\citenamefont
  {Jonnadula}, \citenamefont {Mandayam}, \citenamefont {\ifmmode~\dot{Z}\else
  \.{Z}\fi{}yczkowski},\ and\ \citenamefont
  {Lakshminarayan}}]{PhysRevA.95.040302}%
  \BibitemOpen
  \bibfield  {author} {\bibinfo {author} {\bibfnamefont {B.}~\bibnamefont
  {Jonnadula}}, \bibinfo {author} {\bibfnamefont {P.}~\bibnamefont {Mandayam}},
  \bibinfo {author} {\bibfnamefont {K.}~\bibnamefont {\ifmmode~\dot{Z}\else
  \.{Z}\fi{}yczkowski}},\ and\ \bibinfo {author} {\bibfnamefont
  {A.}~\bibnamefont {Lakshminarayan}},\ }\href
  {https://doi.org/10.1103/PhysRevA.95.040302} {\bibfield  {journal} {\bibinfo
  {journal} {Phys. Rev. A}\ }\textbf {\bibinfo {volume} {95}},\ \bibinfo
  {pages} {040302} (\bibinfo {year} {2017})}\BibitemShut {NoStop}%
\bibitem [{\citenamefont {Clarisse}\ \emph {et~al.}(2007)\citenamefont
  {Clarisse}, \citenamefont {Ghosh}, \citenamefont {Severini},\ and\
  \citenamefont {Sudbery}}]{CLARISSE2007400}%
  \BibitemOpen
  \bibfield  {author} {\bibinfo {author} {\bibfnamefont {L.}~\bibnamefont
  {Clarisse}}, \bibinfo {author} {\bibfnamefont {S.}~\bibnamefont {Ghosh}},
  \bibinfo {author} {\bibfnamefont {S.}~\bibnamefont {Severini}},\ and\
  \bibinfo {author} {\bibfnamefont {A.}~\bibnamefont {Sudbery}},\ }\href
  {https://doi.org/https://doi.org/10.1016/j.physleta.2007.02.001} {\bibfield
  {journal} {\bibinfo  {journal} {Physics Letters A}\ }\textbf {\bibinfo
  {volume} {365}},\ \bibinfo {pages} {400} (\bibinfo {year}
  {2007})}\BibitemShut {NoStop}%
\bibitem [{\citenamefont {Zanardi}(2001)}]{PhysRevA.63.040304}%
  \BibitemOpen
  \bibfield  {author} {\bibinfo {author} {\bibfnamefont {P.}~\bibnamefont
  {Zanardi}},\ }\href {https://doi.org/10.1103/PhysRevA.63.040304} {\bibfield
  {journal} {\bibinfo  {journal} {Phys. Rev. A}\ }\textbf {\bibinfo {volume}
  {63}},\ \bibinfo {pages} {040304} (\bibinfo {year} {2001})}\BibitemShut
  {NoStop}%
\bibitem [{\citenamefont {Rezakhani}(2004)}]{PhysRevA.70.052313}%
  \BibitemOpen
  \bibfield  {author} {\bibinfo {author} {\bibfnamefont {A.~T.}\ \bibnamefont
  {Rezakhani}},\ }\href {https://doi.org/10.1103/PhysRevA.70.052313} {\bibfield
   {journal} {\bibinfo  {journal} {Phys. Rev. A}\ }\textbf {\bibinfo {volume}
  {70}},\ \bibinfo {pages} {052313} (\bibinfo {year} {2004})}\BibitemShut
  {NoStop}%
\bibitem [{\citenamefont {Balakrishnan}\ and\ \citenamefont
  {Sankaranarayanan}(2009)}]{PhysRevA.79.052339}%
  \BibitemOpen
  \bibfield  {author} {\bibinfo {author} {\bibfnamefont {S.}~\bibnamefont
  {Balakrishnan}}\ and\ \bibinfo {author} {\bibfnamefont {R.}~\bibnamefont
  {Sankaranarayanan}},\ }\href {https://doi.org/10.1103/PhysRevA.79.052339}
  {\bibfield  {journal} {\bibinfo  {journal} {Phys. Rev. A}\ }\textbf {\bibinfo
  {volume} {79}},\ \bibinfo {pages} {052339} (\bibinfo {year}
  {2009})}\BibitemShut {NoStop}%
\bibitem [{\citenamefont {Shen}\ and\ \citenamefont {Chen}(2018)}]{Shen_2018}%
  \BibitemOpen
  \bibfield  {author} {\bibinfo {author} {\bibfnamefont {Y.}~\bibnamefont
  {Shen}}\ and\ \bibinfo {author} {\bibfnamefont {L.}~\bibnamefont {Chen}},\
  }\href {https://doi.org/10.1088/1751-8121/aad7cb} {\bibfield  {journal}
  {\bibinfo  {journal} {Journal of Physics A: Mathematical and Theoretical}\
  }\textbf {\bibinfo {volume} {51}},\ \bibinfo {pages} {395303} (\bibinfo
  {year} {2018})}\BibitemShut {NoStop}%
\bibitem [{\citenamefont {Wang}\ \emph {et~al.}(2003)\citenamefont {Wang},
  \citenamefont {Sanders},\ and\ \citenamefont {Berry}}]{PhysRevA.67.042323}%
  \BibitemOpen
  \bibfield  {author} {\bibinfo {author} {\bibfnamefont {X.}~\bibnamefont
  {Wang}}, \bibinfo {author} {\bibfnamefont {B.~C.}\ \bibnamefont {Sanders}},\
  and\ \bibinfo {author} {\bibfnamefont {D.~W.}\ \bibnamefont {Berry}},\ }\href
  {https://doi.org/10.1103/PhysRevA.67.042323} {\bibfield  {journal} {\bibinfo
  {journal} {Phys. Rev. A}\ }\textbf {\bibinfo {volume} {67}},\ \bibinfo
  {pages} {042323} (\bibinfo {year} {2003})}\BibitemShut {NoStop}%
\bibitem [{\citenamefont {Yang}\ \emph {et~al.}(2008)\citenamefont {Yang},
  \citenamefont {Wang},\ and\ \citenamefont {Sun}}]{YANG20084369}%
  \BibitemOpen
  \bibfield  {author} {\bibinfo {author} {\bibfnamefont {Y.}~\bibnamefont
  {Yang}}, \bibinfo {author} {\bibfnamefont {X.}~\bibnamefont {Wang}},\ and\
  \bibinfo {author} {\bibfnamefont {Z.}~\bibnamefont {Sun}},\ }\href
  {https://doi.org/https://doi.org/10.1016/j.physleta.2008.04.023} {\bibfield
  {journal} {\bibinfo  {journal} {Physics Letters A}\ }\textbf {\bibinfo
  {volume} {372}},\ \bibinfo {pages} {4369} (\bibinfo {year}
  {2008})}\BibitemShut {NoStop}%
\bibitem [{\citenamefont {Musz}\ \emph {et~al.}(2013)\citenamefont {Musz},
  \citenamefont {Ku\ifmmode~\acute{s}\else \'{s}\fi{}},\ and\ \citenamefont
  {\ifmmode~\dot{Z}\else \.{Z}\fi{}yczkowski}}]{PhysRevA.87.022111}%
  \BibitemOpen
  \bibfield  {author} {\bibinfo {author} {\bibfnamefont {M.}~\bibnamefont
  {Musz}}, \bibinfo {author} {\bibfnamefont {M.}~\bibnamefont
  {Ku\ifmmode~\acute{s}\else \'{s}\fi{}}},\ and\ \bibinfo {author}
  {\bibfnamefont {K.}~\bibnamefont {\ifmmode~\dot{Z}\else
  \.{Z}\fi{}yczkowski}},\ }\href {https://doi.org/10.1103/PhysRevA.87.022111}
  {\bibfield  {journal} {\bibinfo  {journal} {Phys. Rev. A}\ }\textbf {\bibinfo
  {volume} {87}},\ \bibinfo {pages} {022111} (\bibinfo {year}
  {2013})}\BibitemShut {NoStop}%
\bibitem [{\citenamefont {Chen}\ \emph {et~al.}(2019)\citenamefont {Chen},
  \citenamefont {Ji}, \citenamefont {Kribs}, \citenamefont {Zeng},\ and\
  \citenamefont {Zhang}}]{Chen_2019}%
  \BibitemOpen
  \bibfield  {author} {\bibinfo {author} {\bibfnamefont {J.}~\bibnamefont
  {Chen}}, \bibinfo {author} {\bibfnamefont {Z.}~\bibnamefont {Ji}}, \bibinfo
  {author} {\bibfnamefont {D.~W.}\ \bibnamefont {Kribs}}, \bibinfo {author}
  {\bibfnamefont {B.}~\bibnamefont {Zeng}},\ and\ \bibinfo {author}
  {\bibfnamefont {F.}~\bibnamefont {Zhang}},\ }\href
  {https://doi.org/10.1088/1751-8121/ab15e3} {\bibfield  {journal} {\bibinfo
  {journal} {Journal of Physics A: Mathematical and Theoretical}\ }\textbf
  {\bibinfo {volume} {52}},\ \bibinfo {pages} {215302} (\bibinfo {year}
  {2019})}\BibitemShut {NoStop}%
\bibitem [{\citenamefont {Kong}\ and\ \citenamefont {Zhao}(2024)}]{Kong_2024}%
  \BibitemOpen
  \bibfield  {author} {\bibinfo {author} {\bibfnamefont {F.-Z.}\ \bibnamefont
  {Kong}}\ and\ \bibinfo {author} {\bibfnamefont {J.-L.}\ \bibnamefont
  {Zhao}},\ }\href {https://doi.org/10.1088/1612-202X/ad3627} {\bibfield
  {journal} {\bibinfo  {journal} {Laser Physics Letters}\ }\textbf {\bibinfo
  {volume} {21}},\ \bibinfo {pages} {055206} (\bibinfo {year}
  {2024})}\BibitemShut {NoStop}%
\bibitem [{\citenamefont {Scott}(2004)}]{PhysRevA.69.052330}%
  \BibitemOpen
  \bibfield  {author} {\bibinfo {author} {\bibfnamefont {A.~J.}\ \bibnamefont
  {Scott}},\ }\href {https://doi.org/10.1103/PhysRevA.69.052330} {\bibfield
  {journal} {\bibinfo  {journal} {Phys. Rev. A}\ }\textbf {\bibinfo {volume}
  {69}},\ \bibinfo {pages} {052330} (\bibinfo {year} {2004})}\BibitemShut
  {NoStop}%
\bibitem [{\citenamefont {Linowski}\ \emph {et~al.}(2020)\citenamefont
  {Linowski}, \citenamefont {Rajchel-Mieldzioć},\ and\ \citenamefont
  {Życzkowski}}]{Linowski_2020}%
  \BibitemOpen
  \bibfield  {author} {\bibinfo {author} {\bibfnamefont {T.}~\bibnamefont
  {Linowski}}, \bibinfo {author} {\bibfnamefont {G.}~\bibnamefont
  {Rajchel-Mieldzioć}},\ and\ \bibinfo {author} {\bibfnamefont
  {K.}~\bibnamefont {Życzkowski}},\ }\href
  {https://doi.org/10.1088/1751-8121/ab749a} {\bibfield  {journal} {\bibinfo
  {journal} {Journal of Physics A: Mathematical and Theoretical}\ }\textbf
  {\bibinfo {volume} {53}},\ \bibinfo {pages} {125303} (\bibinfo {year}
  {2020})}\BibitemShut {NoStop}%
\bibitem [{\citenamefont {Wang}\ and\ \citenamefont
  {Zanardi}(2002)}]{PhysRevA.66.044303}%
  \BibitemOpen
  \bibfield  {author} {\bibinfo {author} {\bibfnamefont {X.}~\bibnamefont
  {Wang}}\ and\ \bibinfo {author} {\bibfnamefont {P.}~\bibnamefont {Zanardi}},\
  }\href {https://doi.org/10.1103/PhysRevA.66.044303} {\bibfield  {journal}
  {\bibinfo  {journal} {Phys. Rev. A}\ }\textbf {\bibinfo {volume} {66}},\
  \bibinfo {pages} {044303} (\bibinfo {year} {2002})}\BibitemShut {NoStop}%
\bibitem [{\citenamefont {Makhlin}(2002)}]{makhlin2002nonlocal}%
  \BibitemOpen
  \bibfield  {author} {\bibinfo {author} {\bibfnamefont {Y.}~\bibnamefont
  {Makhlin}},\ }\href {https://doi.org/10.1023/A:1022144002391} {\bibfield
  {journal} {\bibinfo  {journal} {Quantum Information Processing}\ }\textbf
  {\bibinfo {volume} {1}},\ \bibinfo {pages} {243} (\bibinfo {year}
  {2002})}\BibitemShut {NoStop}%
\bibitem [{\citenamefont {Balakrishnan}\ and\ \citenamefont
  {Sankaranarayanan}(2010)}]{PhysRevA.82.034301}%
  \BibitemOpen
  \bibfield  {author} {\bibinfo {author} {\bibfnamefont {S.}~\bibnamefont
  {Balakrishnan}}\ and\ \bibinfo {author} {\bibfnamefont {R.}~\bibnamefont
  {Sankaranarayanan}},\ }\href {https://doi.org/10.1103/PhysRevA.82.034301}
  {\bibfield  {journal} {\bibinfo  {journal} {Phys. Rev. A}\ }\textbf {\bibinfo
  {volume} {82}},\ \bibinfo {pages} {034301} (\bibinfo {year}
  {2010})}\BibitemShut {NoStop}%
\bibitem [{\citenamefont {Singh}\ and\ \citenamefont
  {Nechita}(2022)}]{Singh_2022}%
  \BibitemOpen
  \bibfield  {author} {\bibinfo {author} {\bibfnamefont {S.}~\bibnamefont
  {Singh}}\ and\ \bibinfo {author} {\bibfnamefont {I.}~\bibnamefont
  {Nechita}},\ }\href {https://doi.org/10.1088/1751-8121/ac7017} {\bibfield
  {journal} {\bibinfo  {journal} {Journal of Physics A: Mathematical and
  Theoretical}\ }\textbf {\bibinfo {volume} {55}},\ \bibinfo {pages} {255302}
  (\bibinfo {year} {2022})}\BibitemShut {NoStop}%
\bibitem [{\citenamefont {Scott}\ and\ \citenamefont
  {Caves}(2003)}]{A_J_Scott_2003}%
  \BibitemOpen
  \bibfield  {author} {\bibinfo {author} {\bibfnamefont {A.~J.}\ \bibnamefont
  {Scott}}\ and\ \bibinfo {author} {\bibfnamefont {C.~M.}\ \bibnamefont
  {Caves}},\ }\href {https://doi.org/10.1088/0305-4470/36/36/308} {\bibfield
  {journal} {\bibinfo  {journal} {Journal of Physics A: Mathematical and
  General}\ }\textbf {\bibinfo {volume} {36}},\ \bibinfo {pages} {9553}
  (\bibinfo {year} {2003})}\BibitemShut {NoStop}%
\bibitem [{\citenamefont {Pal}\ and\ \citenamefont
  {Lakshminarayan}(2018)}]{PhysRevB.98.174304}%
  \BibitemOpen
  \bibfield  {author} {\bibinfo {author} {\bibfnamefont {R.}~\bibnamefont
  {Pal}}\ and\ \bibinfo {author} {\bibfnamefont {A.}~\bibnamefont
  {Lakshminarayan}},\ }\href {https://doi.org/10.1103/PhysRevB.98.174304}
  {\bibfield  {journal} {\bibinfo  {journal} {Phys. Rev. B}\ }\textbf {\bibinfo
  {volume} {98}},\ \bibinfo {pages} {174304} (\bibinfo {year}
  {2018})}\BibitemShut {NoStop}%
\bibitem [{\citenamefont {Aravinda}\ \emph {et~al.}(2021)\citenamefont
  {Aravinda}, \citenamefont {Rather},\ and\ \citenamefont
  {Lakshminarayan}}]{PhysRevResearch.3.043034}%
  \BibitemOpen
  \bibfield  {author} {\bibinfo {author} {\bibfnamefont {S.}~\bibnamefont
  {Aravinda}}, \bibinfo {author} {\bibfnamefont {S.~A.}\ \bibnamefont
  {Rather}},\ and\ \bibinfo {author} {\bibfnamefont {A.}~\bibnamefont
  {Lakshminarayan}},\ }\href {https://doi.org/10.1103/PhysRevResearch.3.043034}
  {\bibfield  {journal} {\bibinfo  {journal} {Phys. Rev. Res.}\ }\textbf
  {\bibinfo {volume} {3}},\ \bibinfo {pages} {043034} (\bibinfo {year}
  {2021})}\BibitemShut {NoStop}%
\bibitem [{\citenamefont {Chryssomalakos}\ \emph {et~al.}(2018)\citenamefont
  {Chryssomalakos}, \citenamefont {Guzm\'an-Gonz\'alez},\ and\ \citenamefont
  {Serrano-Ens\'astiga}}]{Chr.Guz.Ser:18}%
  \BibitemOpen
  \bibfield  {author} {\bibinfo {author} {\bibfnamefont {C.}~\bibnamefont
  {Chryssomalakos}}, \bibinfo {author} {\bibfnamefont {E.}~\bibnamefont
  {Guzm\'an-Gonz\'alez}},\ and\ \bibinfo {author} {\bibfnamefont
  {E.}~\bibnamefont {Serrano-Ens\'astiga}},\ }\href
  {https://doi.org/https://doi.org/10.1088/1751-8121/aab349} {\bibfield
  {journal} {\bibinfo  {journal} {J.{} Phys.{} A: Math.{} Theor.}\ }\textbf
  {\bibinfo {volume} {51}},\ \bibinfo {pages} {165202} (\bibinfo {year}
  {2018})}\BibitemShut {NoStop}%
\bibitem [{\citenamefont {Morachis~Galindo}\ and\ \citenamefont
  {Maytorena}(2022)}]{PhysRevA.105.012601}%
  \BibitemOpen
  \bibfield  {author} {\bibinfo {author} {\bibfnamefont {D.}~\bibnamefont
  {Morachis~Galindo}}\ and\ \bibinfo {author} {\bibfnamefont {J.~A.}\
  \bibnamefont {Maytorena}},\ }\href
  {https://doi.org/10.1103/PhysRevA.105.012601} {\bibfield  {journal} {\bibinfo
   {journal} {Phys. Rev. A}\ }\textbf {\bibinfo {volume} {105}},\ \bibinfo
  {pages} {012601} (\bibinfo {year} {2022})}\BibitemShut {NoStop}%
\bibitem [{\citenamefont {Byrd}(1998)}]{10.1063/1.532618}%
  \BibitemOpen
  \bibfield  {author} {\bibinfo {author} {\bibfnamefont {M.}~\bibnamefont
  {Byrd}},\ }\href {https://doi.org/10.1063/1.532618} {\bibfield  {journal}
  {\bibinfo  {journal} {Journal of Mathematical Physics}\ }\textbf {\bibinfo
  {volume} {39}},\ \bibinfo {pages} {6125} (\bibinfo {year}
  {1998})}\BibitemShut {NoStop}%
\bibitem [{\citenamefont {Denis}\ and\ \citenamefont
  {Martin}(2022)}]{Den.Mar:22}%
  \BibitemOpen
  \bibfield  {author} {\bibinfo {author} {\bibfnamefont {J.}~\bibnamefont
  {Denis}}\ and\ \bibinfo {author} {\bibfnamefont {J.}~\bibnamefont {Martin}},\
  }\href {https://doi.org/10.1103/PhysRevResearch.4.013178} {\bibfield
  {journal} {\bibinfo  {journal} {Phys. Rev. Res.}\ }\textbf {\bibinfo {volume}
  {4}},\ \bibinfo {pages} {013178} (\bibinfo {year} {2022})}\BibitemShut
  {NoStop}%
\bibitem [{\citenamefont {Varshalovich}\ \emph {et~al.}(1988)\citenamefont
  {Varshalovich}, \citenamefont {Moskalev},\ and\ \citenamefont
  {Khersonskii}}]{Var.Mos.Khe:88}%
  \BibitemOpen
  \bibfield  {author} {\bibinfo {author} {\bibfnamefont {D.}~\bibnamefont
  {Varshalovich}}, \bibinfo {author} {\bibfnamefont {A.}~\bibnamefont
  {Moskalev}},\ and\ \bibinfo {author} {\bibfnamefont {V.}~\bibnamefont
  {Khersonskii}},\ }\href@noop {} {\emph {\bibinfo {title} {Quantum {T}heory of
  {A}ngular {M}omentum}}}\ (\bibinfo  {publisher} {World Scientific},\ \bibinfo
  {year} {1988})\BibitemShut {NoStop}%
\bibitem [{Note1()}]{Note1}%
  \BibitemOpen
  \bibinfo {note} {Here, we consider the $Q$ matrix of \cite
  {PhysRevA.105.012601} in the symmetric subspace of $\protect \mathcal
  {H}_{1/2}^{\otimes 2}$.}\BibitemShut {Stop}%
\bibitem [{\citenamefont {Agarwal}(1981)}]{PhysRevA.24.2889}%
  \BibitemOpen
  \bibfield  {author} {\bibinfo {author} {\bibfnamefont {G.~S.}\ \bibnamefont
  {Agarwal}},\ }\href {https://doi.org/10.1103/PhysRevA.24.2889} {\bibfield
  {journal} {\bibinfo  {journal} {Phys. Rev. A}\ }\textbf {\bibinfo {volume}
  {24}},\ \bibinfo {pages} {2889} (\bibinfo {year} {1981})}\BibitemShut
  {NoStop}%
\bibitem [{\citenamefont {Serrano-Ens\'astiga}\ and\ \citenamefont
  {Braun}(2020)}]{Ser.Bra:20}%
  \BibitemOpen
  \bibfield  {author} {\bibinfo {author} {\bibfnamefont {E.}~\bibnamefont
  {Serrano-Ens\'astiga}}\ and\ \bibinfo {author} {\bibfnamefont
  {D.}~\bibnamefont {Braun}},\ }\href
  {https://doi.org/10.1103/PhysRevA.101.022332} {\bibfield  {journal} {\bibinfo
   {journal} {Physical Review A}\ }\textbf {\bibinfo {volume} {101}},\ \bibinfo
  {pages} {022332} (\bibinfo {year} {2020})}\BibitemShut {NoStop}%
\bibitem [{\citenamefont {Baguette}\ and\ \citenamefont
  {Martin}(2017)}]{Bag.Mar:17}%
  \BibitemOpen
  \bibfield  {author} {\bibinfo {author} {\bibfnamefont {D.}~\bibnamefont
  {Baguette}}\ and\ \bibinfo {author} {\bibfnamefont {J.}~\bibnamefont
  {Martin}},\ }\href {https://doi.org/10.1103/PhysRevA.96.032304} {\bibfield
  {journal} {\bibinfo  {journal} {Phys. Rev. A}\ }\textbf {\bibinfo {volume}
  {96}},\ \bibinfo {pages} {032304} (\bibinfo {year} {2017})}\BibitemShut
  {NoStop}%
\bibitem [{Note2()}]{Note2}%
  \BibitemOpen
  \bibinfo {note} {Using the multipole expansion~\protect \eqref
  {rhoTLMexpansion}, each spin-1 operator has associated to it a spin-2 state.
  Specifically, the components $\protect \bm {\rho }_2 (U)$ can be expressed as
  $\mathinner {|{\protect \bm {\rho }_2 (U)}\rangle } = \DOTSB \sum@ \slimits@
  _{m=-2}^2 \rho _{2m}(U) \mathinner {|{2,m}\rangle }$. In particular, for $U=A
  \in \protect \mathrm {SU}(3)$, where $A$ is defined in Eq.~\protect \eqref
  {Ec:Anonlocal}, we obtain that $\mathinner {|{\protect \bm {\rho }_2
  (A^{*2})}\rangle } \propto \mathinner {|{\psi _{A}}\rangle }$ (see
  Eq.~\protect \eqref {Eq.State.A}). However, this proportionality does not
  extend to higher spins.}\BibitemShut {Stop}%
\bibitem [{Note3()}]{Note3}%
  \BibitemOpen
  \bibinfo {note} {The spin-3 states shown in Eq.~\protect \eqref {Eq.Bases.N3}
  span a 3-dimensional 1-anticoherent subspace (see Ref.~\cite {Ser.Chr.Mar:24}
  for more details).}\BibitemShut {Stop}%
\bibitem [{\citenamefont {Martin}\ \emph {et~al.}(2010)\citenamefont {Martin},
  \citenamefont {Giraud}, \citenamefont {Braun}, \citenamefont {Braun},\ and\
  \citenamefont {Bastin}}]{PhysRevA.81.062347}%
  \BibitemOpen
  \bibfield  {author} {\bibinfo {author} {\bibfnamefont {J.}~\bibnamefont
  {Martin}}, \bibinfo {author} {\bibfnamefont {O.}~\bibnamefont {Giraud}},
  \bibinfo {author} {\bibfnamefont {P.~A.}\ \bibnamefont {Braun}}, \bibinfo
  {author} {\bibfnamefont {D.}~\bibnamefont {Braun}},\ and\ \bibinfo {author}
  {\bibfnamefont {T.}~\bibnamefont {Bastin}},\ }\href
  {https://doi.org/10.1103/PhysRevA.81.062347} {\bibfield  {journal} {\bibinfo
  {journal} {Phys. Rev. A}\ }\textbf {\bibinfo {volume} {81}},\ \bibinfo
  {pages} {062347} (\bibinfo {year} {2010})}\BibitemShut {NoStop}%
\bibitem [{\citenamefont {Giraud}\ \emph {et~al.}(2010)\citenamefont {Giraud},
  \citenamefont {Braun},\ and\ \citenamefont {Braun}}]{Gir.Bra.Bra:10}%
  \BibitemOpen
  \bibfield  {author} {\bibinfo {author} {\bibfnamefont {O.}~\bibnamefont
  {Giraud}}, \bibinfo {author} {\bibfnamefont {P.}~\bibnamefont {Braun}},\ and\
  \bibinfo {author} {\bibfnamefont {D.}~\bibnamefont {Braun}},\ }\href
  {https://doi.org/10.1088/1367-2630/12/6/063005} {\bibfield  {journal}
  {\bibinfo  {journal} {New Journal of Physics}\ }\textbf {\bibinfo {volume}
  {12}},\ \bibinfo {pages} {063005} (\bibinfo {year} {2010})}\BibitemShut
  {NoStop}%
\bibitem [{\citenamefont {Chryssomalakos}\ \emph {et~al.}(2021)\citenamefont
  {Chryssomalakos}, \citenamefont {Hanotel}, \citenamefont
  {Guzm\'an-Gonz\'alez}, \citenamefont {Braun}, \citenamefont
  {Serrano-Ens\'astiga},\ and\ \citenamefont {\.Zyczkowski}}]{Chr.etal:21}%
  \BibitemOpen
  \bibfield  {author} {\bibinfo {author} {\bibfnamefont {C.}~\bibnamefont
  {Chryssomalakos}}, \bibinfo {author} {\bibfnamefont {L.}~\bibnamefont
  {Hanotel}}, \bibinfo {author} {\bibfnamefont {E.}~\bibnamefont
  {Guzm\'an-Gonz\'alez}}, \bibinfo {author} {\bibfnamefont {D.}~\bibnamefont
  {Braun}}, \bibinfo {author} {\bibfnamefont {E.}~\bibnamefont
  {Serrano-Ens\'astiga}},\ and\ \bibinfo {author} {\bibfnamefont
  {K.}~\bibnamefont {\.Zyczkowski}},\ }\href
  {https://doi.org/10.1103/PhysRevA.104.012407} {\bibfield  {journal} {\bibinfo
   {journal} {Phys.{} Rev.{} A}\ }\textbf {\bibinfo {volume} {104}},\ \bibinfo
  {pages} {012407} (\bibinfo {year} {2021})}\BibitemShut {NoStop}%
\bibitem [{\citenamefont {Goldberg}\ \emph {et~al.}(2020)\citenamefont
  {Goldberg}, \citenamefont {Klimov}, \citenamefont {Grassl}, \citenamefont
  {Leuchs},\ and\ \citenamefont {Sánchez-Soto}}]{Gol.Kli.Gra.Leu.San:20}%
  \BibitemOpen
  \bibfield  {author} {\bibinfo {author} {\bibfnamefont {A.~Z.}\ \bibnamefont
  {Goldberg}}, \bibinfo {author} {\bibfnamefont {A.~B.}\ \bibnamefont
  {Klimov}}, \bibinfo {author} {\bibfnamefont {M.}~\bibnamefont {Grassl}},
  \bibinfo {author} {\bibfnamefont {G.}~\bibnamefont {Leuchs}},\ and\ \bibinfo
  {author} {\bibfnamefont {L.~L.}\ \bibnamefont {Sánchez-Soto}},\ }\href
  {https://doi.org/10.1116/5.0025819} {\bibfield  {journal} {\bibinfo
  {journal} {AVS Quantum Science}\ }\textbf {\bibinfo {volume} {2}},\ \bibinfo
  {pages} {044701} (\bibinfo {year} {2020})}\BibitemShut {NoStop}%
\bibitem [{\citenamefont {Zimba}(2006)}]{Zim:06}%
  \BibitemOpen
  \bibfield  {author} {\bibinfo {author} {\bibfnamefont {J.}~\bibnamefont
  {Zimba}},\ }\href@noop {} {\bibfield  {journal} {\bibinfo  {journal}
  {Electronic Journal of Theoretical Physics}\ }\textbf {\bibinfo {volume}
  {3}},\ \bibinfo {pages} {143} (\bibinfo {year} {2006})}\BibitemShut {NoStop}%
\bibitem [{\citenamefont {Crann}\ \emph {et~al.}(2010)\citenamefont {Crann},
  \citenamefont {Pereira},\ and\ \citenamefont {Kribs}}]{Crann_2010}%
  \BibitemOpen
  \bibfield  {author} {\bibinfo {author} {\bibfnamefont {J.}~\bibnamefont
  {Crann}}, \bibinfo {author} {\bibfnamefont {R.}~\bibnamefont {Pereira}},\
  and\ \bibinfo {author} {\bibfnamefont {D.~W.}\ \bibnamefont {Kribs}},\ }\href
  {https://doi.org/10.1088/1751-8113/43/25/255307} {\bibfield  {journal}
  {\bibinfo  {journal} {Journal of Physics A: Mathematical and Theoretical}\
  }\textbf {\bibinfo {volume} {43}},\ \bibinfo {pages} {255307} (\bibinfo
  {year} {2010})}\BibitemShut {NoStop}%
\bibitem [{\citenamefont {Giraud}\ \emph {et~al.}(2015)\citenamefont {Giraud},
  \citenamefont {Braun}, \citenamefont {Baguette}, \citenamefont {Bastin},\
  and\ \citenamefont {Martin}}]{PhysRevLett.114.080401}%
  \BibitemOpen
  \bibfield  {author} {\bibinfo {author} {\bibfnamefont {O.}~\bibnamefont
  {Giraud}}, \bibinfo {author} {\bibfnamefont {D.}~\bibnamefont {Braun}},
  \bibinfo {author} {\bibfnamefont {D.}~\bibnamefont {Baguette}}, \bibinfo
  {author} {\bibfnamefont {T.}~\bibnamefont {Bastin}},\ and\ \bibinfo {author}
  {\bibfnamefont {J.}~\bibnamefont {Martin}},\ }\href
  {https://doi.org/10.1103/PhysRevLett.114.080401} {\bibfield  {journal}
  {\bibinfo  {journal} {Phys. Rev. Lett.}\ }\textbf {\bibinfo {volume} {114}},\
  \bibinfo {pages} {080401} (\bibinfo {year} {2015})}\BibitemShut {NoStop}%
\bibitem [{\citenamefont {Baguette}\ \emph {et~al.}(2015)\citenamefont
  {Baguette}, \citenamefont {Damanet}, \citenamefont {Giraud},\ and\
  \citenamefont {Martin}}]{Bag.Dam.Gir.Mar:15}%
  \BibitemOpen
  \bibfield  {author} {\bibinfo {author} {\bibfnamefont {D.}~\bibnamefont
  {Baguette}}, \bibinfo {author} {\bibfnamefont {F.}~\bibnamefont {Damanet}},
  \bibinfo {author} {\bibfnamefont {O.}~\bibnamefont {Giraud}},\ and\ \bibinfo
  {author} {\bibfnamefont {J.}~\bibnamefont {Martin}},\ }\href
  {https://doi.org/10.1103/PhysRevA.92.052333} {\bibfield  {journal} {\bibinfo
  {journal} {Phys.{} Rev.{} A}\ }\textbf {\bibinfo {volume} {92}},\ \bibinfo
  {eid} {052333} (\bibinfo {year} {2015})}\BibitemShut {NoStop}%
\bibitem [{\citenamefont {Chryssomalakos}\ and\ \citenamefont
  {Hern\'andez-Coronado}(2017)}]{Chr.Her:17}%
  \BibitemOpen
  \bibfield  {author} {\bibinfo {author} {\bibfnamefont {C.}~\bibnamefont
  {Chryssomalakos}}\ and\ \bibinfo {author} {\bibfnamefont {H.}~\bibnamefont
  {Hern\'andez-Coronado}},\ }\href {https://doi.org/10.1103/PhysRevA.95.052125}
  {\bibfield  {journal} {\bibinfo  {journal} {Phys.{} Rev.{} A}\ }\textbf
  {\bibinfo {volume} {95}} (\bibinfo {year} {2017})}\BibitemShut {NoStop}%
\bibitem [{\citenamefont {Goldberg}\ and\ \citenamefont
  {James}(2018)}]{PhysRevA.98.032113}%
  \BibitemOpen
  \bibfield  {author} {\bibinfo {author} {\bibfnamefont {A.~Z.}\ \bibnamefont
  {Goldberg}}\ and\ \bibinfo {author} {\bibfnamefont {D.~F.~V.}\ \bibnamefont
  {James}},\ }\href {https://doi.org/10.1103/PhysRevA.98.032113} {\bibfield
  {journal} {\bibinfo  {journal} {Phys. Rev. A}\ }\textbf {\bibinfo {volume}
  {98}},\ \bibinfo {pages} {032113} (\bibinfo {year} {2018})}\BibitemShut
  {NoStop}%
\bibitem [{\citenamefont {Serrano-Ensástiga}\ \emph
  {et~al.}(2024)\citenamefont {Serrano-Ensástiga}, \citenamefont
  {Chryssomalakos},\ and\ \citenamefont {Martin}}]{Ser.Chr.Mar:24}%
  \BibitemOpen
  \bibfield  {author} {\bibinfo {author} {\bibfnamefont {E.}~\bibnamefont
  {Serrano-Ensástiga}}, \bibinfo {author} {\bibfnamefont {C.}~\bibnamefont
  {Chryssomalakos}},\ and\ \bibinfo {author} {\bibfnamefont {J.}~\bibnamefont
  {Martin}},\ }\href {https://arxiv.org/abs/2404.15548} {\bibinfo {title}
  {Quantum metrology of rotations with mixed spin states}} (\bibinfo {year}
  {2024}),\ \Eprint {https://arxiv.org/abs/2404.15548} {arXiv:2404.15548}
  \BibitemShut {NoStop}%
\bibitem [{\citenamefont {Fano}(1953)}]{Fan:53}%
  \BibitemOpen
  \bibfield  {author} {\bibinfo {author} {\bibfnamefont {U.}~\bibnamefont
  {Fano}},\ }\href {https://doi.org/10.1103/physrev.90.577} {\bibfield
  {journal} {\bibinfo  {journal} {Phys. Rev.}\ }\textbf {\bibinfo {volume}
  {90}},\ \bibinfo {pages} {577} (\bibinfo {year} {1953})}\BibitemShut
  {NoStop}%
\bibitem [{\citenamefont {Zhang}(2015)}]{zhang2014matrix}%
  \BibitemOpen
  \bibfield  {author} {\bibinfo {author} {\bibfnamefont {L.}~\bibnamefont
  {Zhang}},\ }\href {https://arxiv.org/abs/1408.3782} {\bibinfo {title} {Matrix
  integrals over unitary groups: An application of schur-weyl duality}}
  (\bibinfo {year} {2015}),\ \Eprint {https://arxiv.org/abs/1408.3782}
  {arXiv:1408.3782} \BibitemShut {NoStop}%
\end{thebibliography}%
\end{document}